\documentclass[
reprint,
 amsmath,amssymb,
 aip,
]{revtex4-2}

\usepackage{graphicx}
\usepackage{dcolumn}
\usepackage{bm}
\usepackage{color}

\begin{document}

\title{%
Numerical analysis of voltage-controlled magnetization switching operation 
in magnetic-topological-insulator-based devices%
}
\author{Takashi Komine}
\email{takashi.komine.nfm@vc.ibaraki.ac.jp}
\affiliation{%
Graduate School of Science and Engineering, Ibaraki University, 
4-12-1, Nakanarusawa, Hitachi, Ibaraki, 316-8511, Japan}
\author{Takahiro Chiba}
\affiliation{%
Frontier research Institute for Interdisciplinary Sciences,
Tohoku University, Sendai, Miyagi 980-0845, Japan}
\affiliation{%
Department of Applied Physics, Graduate School of Engineering,
Tohoku University, Sendai, Miyagi 980-8579, Japan
}

\begin{abstract}
 We theoretically investigate influences of electronic
 circuit delay, noise and temperature on write-error-rate (WER)
 in voltage-controlled magnetization switching operation of a 
 magnetic-topological-insulator-based (MTI) device by means of the
 micromagnetic simulation.
 This device realizes magnetization switching via spin-orbit torque
 (SOT) and voltage-controlled magnetic anisotropy (VCMA) which
 originate from 2D-Dirac electronic structure.
 We reveal that the device operation is extremely robust against circuit
 delay and signal-to-noise ratio.
 We demonstrate that the WER on the order of approximately $10^{-4}$ or
 below is achieved around room temperature due to steep change in VCMA.
 Also, we show that the larger SOT improves thermal stability factor.
 This study provides a next perspective for developing 
 voltage-driven spintronic devices with ultra-low power consumption.
\end{abstract}


\date{\today}

\maketitle


Electrical control of magnetization has attracted much attention
for next generation spintronic devices such as non-volatile magnetic
memory\cite{Ando_JAP_2014}, high-speed logic\cite{Barla_JCompElec_2021},
and low-power data transmission\cite{Zhang_IEEE_6800517}.
Voltage-controlled magnetization anisotropy (VCMA) is a
promising way to drive magnetization switching with high energy
efficiency, and to realize magnetic memories and logic devices with low
power consumption\cite{Maruyama_NatNano_2009,Endo_APL_2010,
Shiota_NatMater_2011, Shiota_APL_2012, 
Shiota_APEX_2016, Grezes_APL_2016}.
Although VCMA is expected to have superior energy efficiency, the
VCMA-driven magnetization switching faces practical issues including the
write-error rate (WER), narrow operating window, and necessity for
external bias magnetic field\cite{VC-MTJ-review}.
The reduction of WER in VCMA-driven magnetic devices is especially a
serious issue for the practical applications\cite{Shiota_APL_2017,
Yamamoto_PRA_2019}.

While the voltage-control of ferromagnetic metal/non-magnetic insulator
bilayers, such as CoFe/MgO, is commonly studied, a perpendicular voltage
also shifts the Fermi energy of Dirac electrons on the surface of
magnetic topological insulator (MTI). 
Three-dimensional topological insulators (TIs), such as ${\rm
Bi_2Se_3}$, have an insulating bulk and conducting surface
states\cite{Ando_2013}.
MTIs are ferromagnetically ordered by injection of magnetic dopants into
TIs\cite{Tokura_2019}.
Recently, the electric field effect in a magnetic topological insulator
(MTI) and a TI/ferromagnetic insulator bilayer has been investigated by
both theoretical and experimental
studies\cite{Wang_PRL_2015,Sekine_PRB_2016,Semenov_PRB_2012, Fan_2016}.
On the other hand, current-induced spin-orbit torque
(SOT) on TI is another key method of manipulating magnetic moment
\cite{Fan_2016,Pham_Nature_2018}, enabling a deterministic magnetization
switching without external magnetic field
\cite{nmatCai2017,admatCao2020,adeleBekele2021}. Furthermore,
engineering the SOT efficiency by means of a voltage (or an
electric-field) is a crucial method for practical device applications
\cite{nmatCai2017,PRLFilianina2020}. Hence, it is highly desirable to
simultaneously control the magnetic anisotropy
and SOT by a voltage in MTI-based devices for high energy efficiency.

Recently, Chiba {\it et al.} proposed the field-effect-transistor
(FET)-like devices which consists of TI and
MTI\cite{Chiba_PRB_2017,Chiba-PRA2020,Chiba-APL2021}.
They presented two distinct methods for the magnetization switching by
using electric field control of SOT and perpendicular magnetic
anisotropy in TI and MTI hybrid systems.
It was reported clear magnetization switching by combining
adequate source-drain voltage and gate pulse.
The writing energy of $< 0.1{\rm fJ/bit}$ is expected for the practical
use of voltage-induced magnetization switching devices.
The power consumption of the MTI device can be achieved about $1 {\rm
fJ/bit}$, which is sufficiently reduced.
Although the gate pulse shape and the signal-to-noise ratio affect the
WER by peripheral electrical circuits in the practical application, WER
in the MTI device has not been discussed yet.
In this study, we demonstrate magnetization switching in the MTI devices
under external disturbance such as electrical circuit noise and thermal
fluctuation by the micromagnetic simulation.


\begin{figure}[htp] 
 \begin{center}
  \includegraphics[width=7.7truecm,clip]{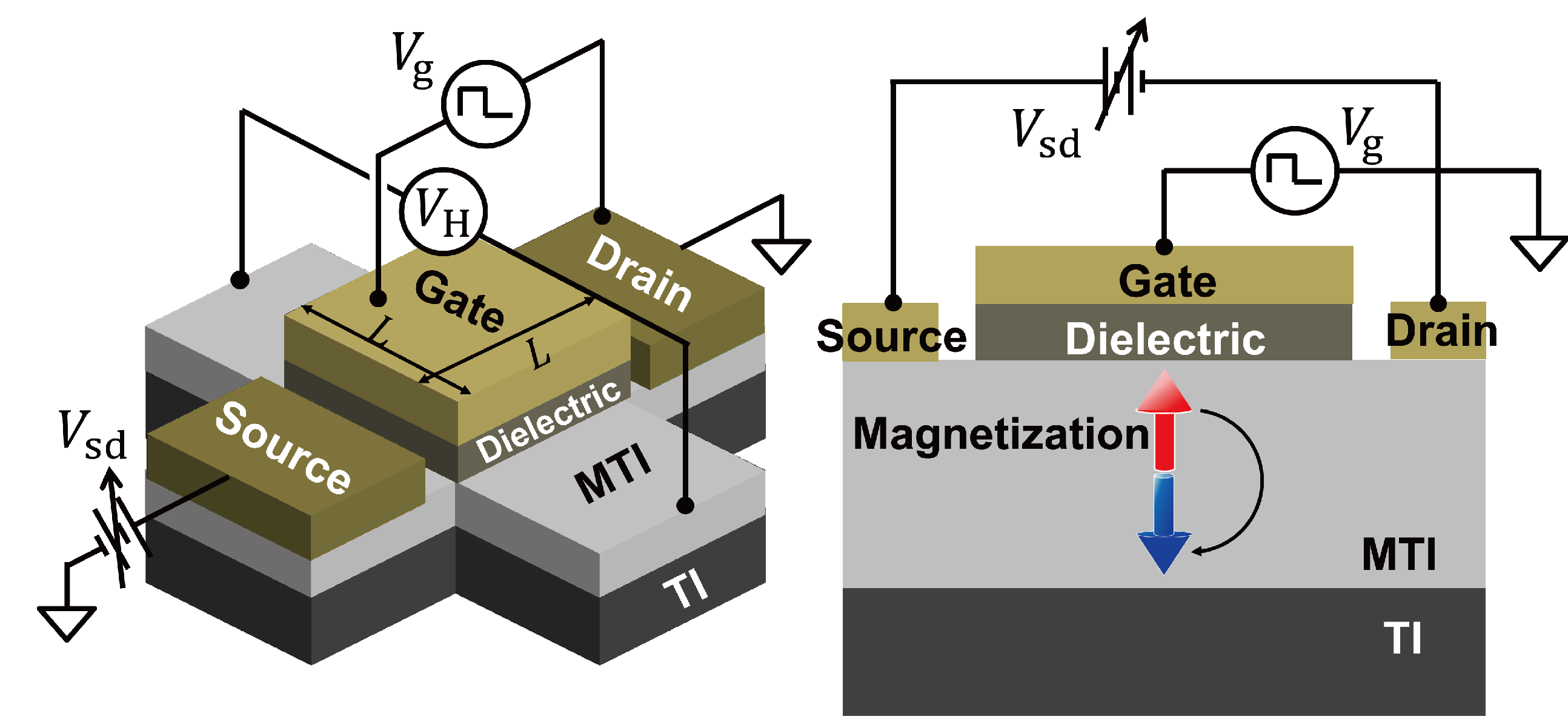}
 \end{center}
 \caption{Schematic illustration of a MTI-based device with a FET-like
 structure consisting of magnetic-TI (MTI) and TI film.}
 \label{fig:device}
\end{figure}

Figure \ref{fig:device} illustrates a MTI-based device involving a MTI
film as in which the top of the surface of MTI is a conduction-channel
layer.
By applying a source-drain electric field $\mathbf{E}_{\rm sd}$ and
gate voltage $V_{g}$, this device realizes magnetization switching via
SOT and VCMA without external bias magnetic field. In this device, the
anomalous Hall effect is used for readout.

In order to analyze the detail operation and WER of the MTI-based device,
the macro-spin model based on the following Landau-Lifshitz-Gilbert(LLG)
equation was utilized:
\begin{equation}
 \dot{\mathbf{m}} =
  \displaystyle
  -\gamma \mathbf{m} \times \mathbf{B}_{\rm eff}
  + \alpha \mathbf{m} \times \dot{\mathbf{m}}
  + \mathbf{T}_{\rm SO}(E_{\rm sd},V_{g}),
\end{equation}
where $\gamma$ is the gyromagnetic constant, $\alpha$ the Gilbert
damping parameter, and the effective field $\mathbf{B}_{\rm eff}$
includes
\begin{equation}
 \mathbf{B}_{\rm eff} =
  \mathbf{B}_{d} + \mathbf{B}_{k}(V_{g}) + \mathbf{B}_{\rm th}(T).
\end{equation}
where $\mathbf{B}_{d}, \mathbf{B}_{k}$ and $\mathbf{B}_{\rm th}$ are the
demagnetizing field, the anisotropy field, and the thermal fluctuation
field, respectively.
The anisotropy field $\mathbf{B}_{k}$ is tunable by the gate voltage
$V_{g}$ or the corresponding Fermi level $E_{F}$.
Here, the SOT $\mathbf{T}_{\rm SO}$ depends on the gate voltage $V_{g}$ and
 source-drain electric field $E_{\rm sd}$,
which are derived from 2D-Dirac electronic structure at the interface
between the dielectric and MTI film\cite{Chiba-PRA2020}.
Moreover, the thermal fluctuation field $\mathbf{B}_{\rm th}$ with
3D Gaussian distribution of dispersion is taken into account. 
\begin{equation}
 \sigma^2 =
  \frac{2 \alpha k_{B} T}{\gamma M_{s} ( 1+\alpha^2 ) V \Delta t},
\end{equation} 

The SOT term $\mathbf{T}_{\rm SO}$ depends on the source-drain electric
field $E_{\rm sd}$ and the gate voltage $V_{g}$ as follow:
\begin{equation}
 \mathbf{T}_{\rm SO}(E_{\rm sd},V_{g})
  = -\frac{\gamma\Delta}{M_{s} d}
  \mathbf{m} \times \bm{\mu}(E_{\rm sd},V_{g})
  \equiv -\gamma \mathbf{m} \times \mathbf{B}_{\rm SOT},
\end{equation}
where $\bm{\mu}$ is the electrically induced nonequilibrium spin
polarization, and $\Delta$ is the spin splitting of 2D-Dirac electrons
or the exchange interaction coupled to the homogeneous localized moment
of MTI.
This term acts the effective fields along the longitudinal component
$\mu_{x}$ and the transverse component $\mu_{y}$ as functions of 
$E_{\rm sd}$ and $V_{g}$ as follows
\begin{eqnarray}
 \bm{\mu}(E_{\rm sd},V_{g}) &=&
  \mu_x(E_{\rm sd},V_{g}) \hat{\mathbf{x}} + \mu_y(E_{\rm sd},V_{g})
  \hat{\mathbf{y}}, \\
  \displaystyle
  \mu_x(E_{\rm sd},V_{g}) &=&
  \frac{4e E_{\rm sd}}{hv_{F}} \Delta
  \frac{E_{F}(V_g) m_z \left[ E^2_F(V_g) + \Delta^2 m^2_z\right]}%
  {\left[ E^2_F(V_g) + 3\Delta^2 m^2_z \right]^2}, \\
  \mu_y(E_{\rm sd},V_{g}) &=&
  \frac{e E_{\rm sd}}{2hv_{F}} \frac{E_{F}(V_g)\tau}{\hbar}
  \frac{E^2_F(V_g)-\Delta^2 m^2_z}{E^2_F(V_g) + 3\Delta^2 m^2_z}.
\end{eqnarray}
Note that the estimated domain size and the domain wall width are about
400~nm and 150~nm, respectively.
The uniform domain appears in the MTI-based device consisting of uniform
materials when the device size is less than 1~$\mu$ m,
and the magnetization switching occurs corresponding to rotation mode 
under the application of gate voltage and source-drain electric field.
Thus, the macro-spin model is valid for analysis of magnetization
switching behavior and WER.

In magnetic tunnel junction (MTJ) devices, the VCMA as a function of
pulse voltage $V_{\rm pulse}$ is a odd function due to charge
accumulation at the magnetic
layer\cite{Weisheit_Science_2007,
Tsujikawa_PRL_2009,Nakamura_PRL_2009,Duan_PRL_2008}.
The VCMA coefficient $\xi$ in the MTI device approximately reaches
$-350~{\rm fJ/Vm}$ estimated as defined $\xi = ( \delta K_{u} t ) /
\delta E_{\rm pulse}$ by the same manner of MTJ\cite{VC-MTJ-review}.
On the other hand, in the MTI-based devices, the VCMA as a function of
gate pulse voltage $V_{g}$ is an even function due to the Fermi level
tuning in the gap-opened Dirac dispersion.
By the same manner, the VCMA coefficient is about $-4~{\rm fJ/Vm}$ but
the anisotropy $K_{u}$ reaches to zero at the gate voltage of $0.4~{\rm
V}$.
In the MTI-based device, comparably small gate voltage can control the
maximum to the minimum in the Dirac magnetic anisotropy $K_{u}$.

In MTJ, the VCMA effect has been used to assist the magnetization
switching in combination with an external
field\cite{Shiota_APEX_2009,Amiri_JAP_2013,Han_IEEE_2015}, 
the spin-transfer-torque\cite{Kanai_APL_2014}, 
the crystallographic strain\cite{Kato_2018},
thermally excitation\cite{Yamamoto_PRA_2018}, 
or the SOT\cite{Yoda_IEDM_2016, Inokuchi_2017,
Baek_NatElec_2018,Mishra_NatComm_2019} to reduce the write energy.
In the MTI-based device, the SOTs including both field-like and
damping-like terms assists the VCMA magnetization switching by applying
the source-drain electric field $E_{sd}$.

\begin{figure}[htp!] 
 \begin{center}
  \includegraphics[width=5.3truecm,angle=270]{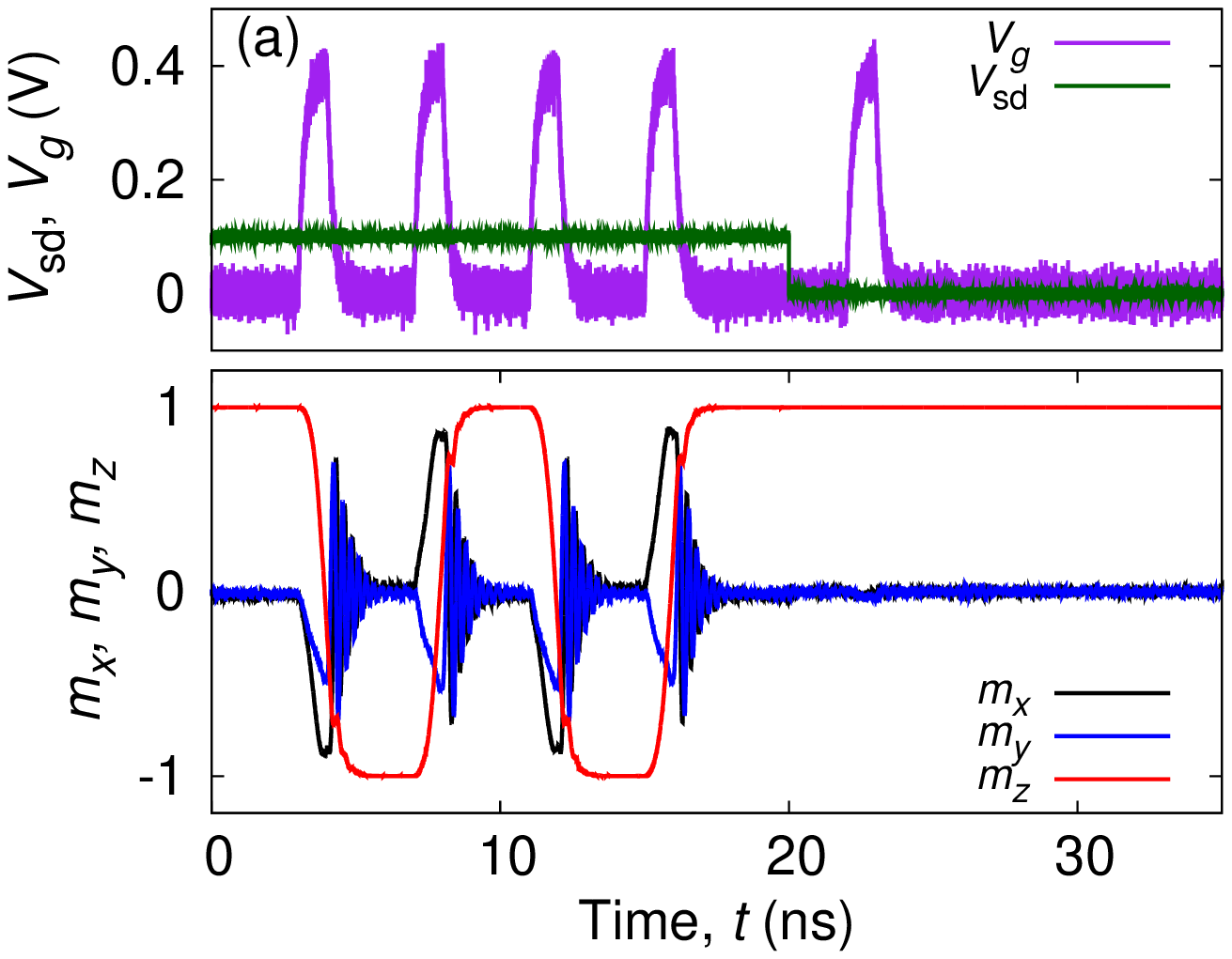}
  \includegraphics[width=6.3truecm,angle=270]{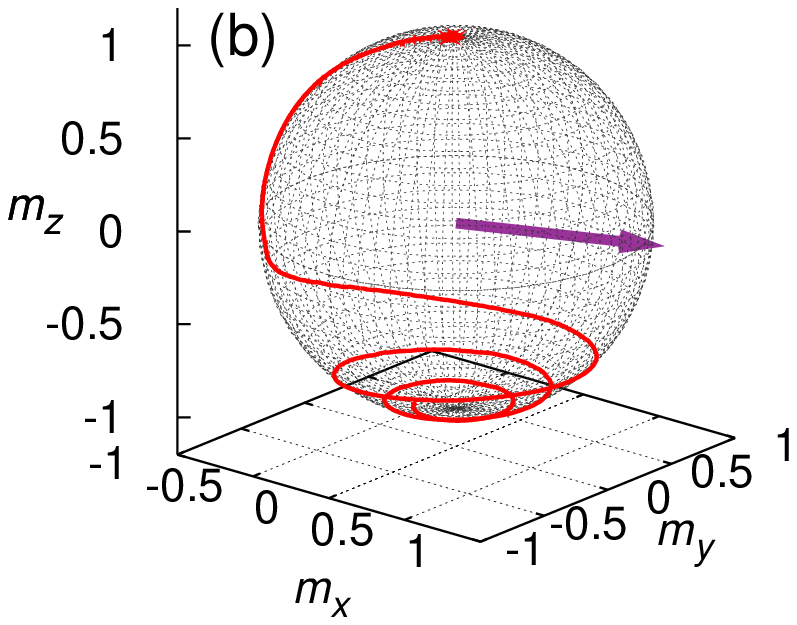}
  \includegraphics[width=6.3truecm,angle=270]{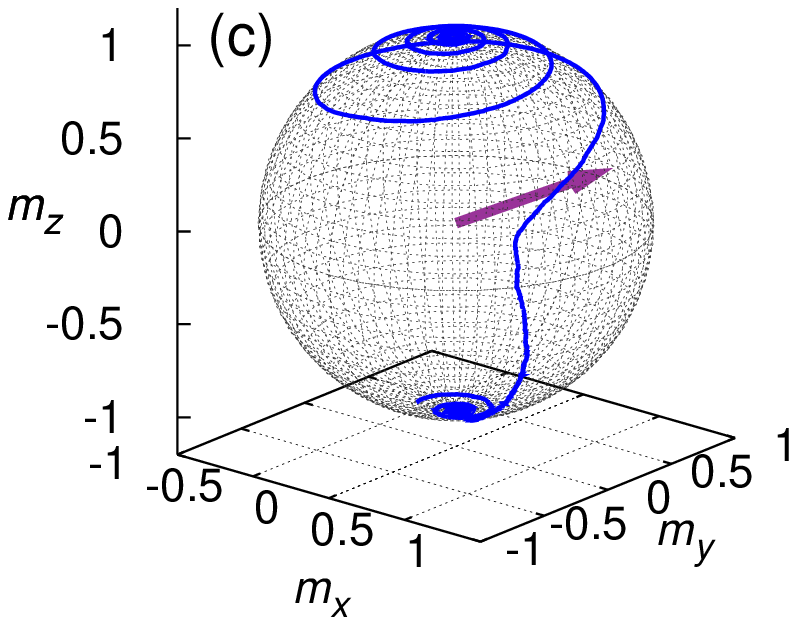}
 \end{center}
 \caption{(a) Magnetization switching behavior under the application of
  source-drain electric field $E_{\rm sd}$ of $1 \times 10^{5}~{\rm
 V\,m^{-1}}$ and gate pulse with the duration $\tau_{g}$ of $1~{\rm ns}$
 and the gate voltage $V_{g}$ of $0.39~{\rm V}$.
 The gate pulse is passed through the 1st-order filter with $f_{c}$ of
 $5~{\rm GHz}$, the SNRs for both gate and source-drain voltages are
 $27~{\rm dB}$, and the temperature $T$ is set to 300~{\rm K}.
 Corresponding magnetization trajectories of (b) changing from upward to
 downward and (c) changing from upward to downward.
 The purple arrow indicate an effective magnetic field due to the
 spin-orbit torque.
 }
 \label{fig:waveform}
\end{figure}

Although in the previous report, the ideal square wave of gate pulse
with pulse duration $\tau_{g}$ are applied for magnetization switching,
the practical pulse has the electronic circuit delay and electrical
noise.
In this study, the 1st-order filter with cut-off frequency $f_{c}$,
\begin{equation}
 G(f) = \frac{1}{1+j \frac{f}{f_c}},
\end{equation}
was used as the circuit delay, and the cut-off dependence of WER
was investigated.
The device has dimensions of $L~\times~L~\times~7~{\rm nm}$,
and the length $L$ is $1$~{$\mu$ m} as the standard size in the
simulation.
The material parameters related to MTI-based system are listed in Table
\ref{tbl:mater}.

\begin{table}
 \caption{Material parameters.}
 \label{tbl:mater}
 \centering
 \begin{tabular}{cc}
  \hline
  Gilbert damping, $\alpha$ & $0.1$ \\
  Surface gap in Dirac cone, $2\Delta~{\rm (meV)}$
  \cite{Hirahara_NanoLett_2017,Tokura_2019,Mogi_PRL_2019}
  & $60$ \\
  Bulk band gap, $2\Delta_{c}~{\rm (meV)}$ & $200$ \\
  Fermi velocity in Dirac cone, $v_{F}~{\rm (m\,s^{-1})}$ &
      $4 \times 10^5$ \\
  Dielectric constant in TI, $\epsilon_{r}$ & $9.7$ \\
  Saturation magnetization, $M_{s}~{\rm
  (kA\,m^{-1})}$\cite{Fan_NatNano_2016} & $8.5$ \\
  TI thickness, $d_{\rm TI}~{\rm (nm)}$ & $7$ \\
  Dielectric thicknesses, $d_{D}~{\rm (nm)}$ & $20$ \\
  \hline
 \end{tabular}
\end{table}

At the beginning, we demonstrate magnetization reversal in the MTI-based
device under the circumstance with typical circuit and thermal
noises.
In the MTI-based device, the VCMA is a trigger to switch the
magnetization direction while the SOT induces magnetization switching
trajectory, {\it i.e.}, the SOT due to source-drain electric field takes
a role of an external magnetic field in VCMA-MTJ devices.

Figure \ref{fig:waveform} shows magnetization switching behavior under
the application of source-drain and gate voltages.
When the gate voltage $V_{g}$ of $0.39~{\rm V}$ is applied, the
magnetic anisotropy sharply goes down to zero, enabling the
magnetization to be along an arbitrary direction 
(see also Supplementary material).
Hence, during the duration of $V_{g}$ the SOT due to $E_{\rm sd}$
derives to magnetization reversal.
On the other hand, when $V_{\rm sd}$ is not applied, the magnetization
is not reversed even if the gate pulse is applied.
Corresponding energy distribution as a function of azimuth angle and
$z$-component of magnetization are shown in the 
Supplementary material.
Such 
a feature of this device is that magnetization reversal occurs
by applying an appropriate gate pulse regardless of the polarity of the
magnetization from upward to downward or from downward to upward.

\begin{figure}[htp] 
 \begin{center}
  \includegraphics[width=5.9truecm]{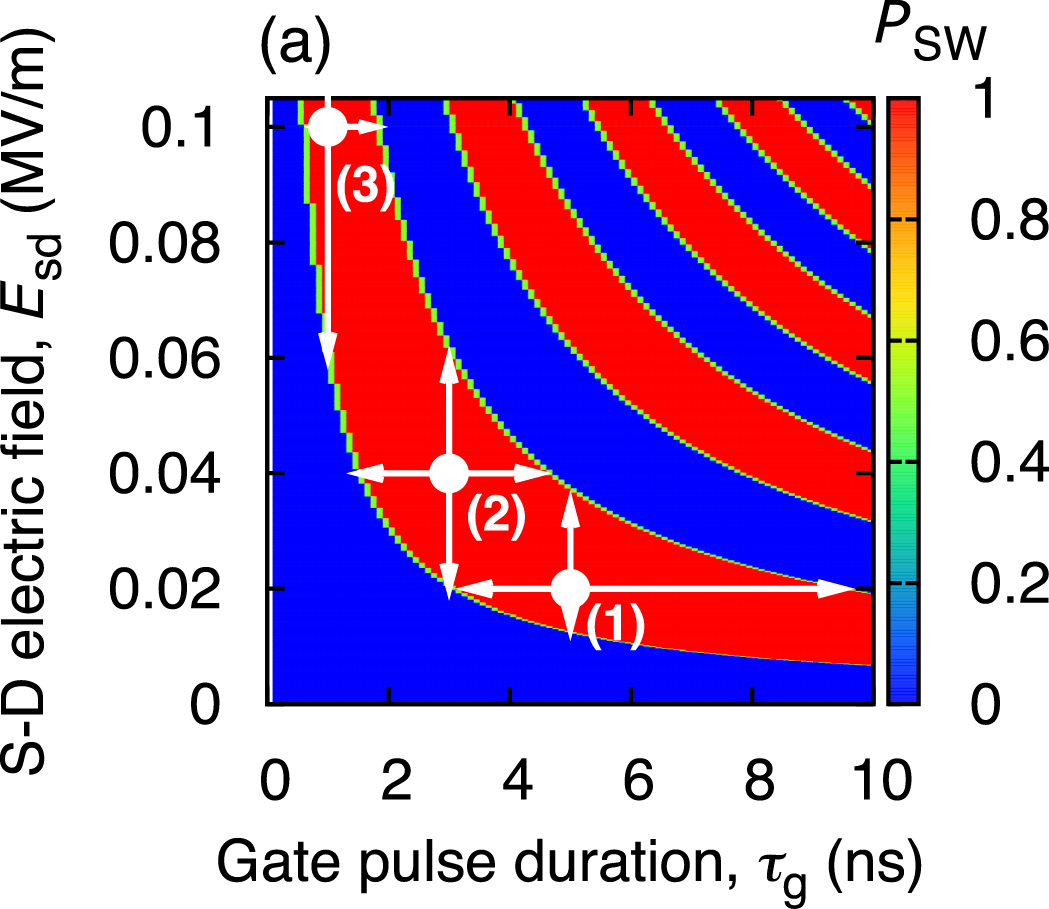}
 \end{center}
 \begin{center}
  \includegraphics[width=5.3truecm,angle=270]{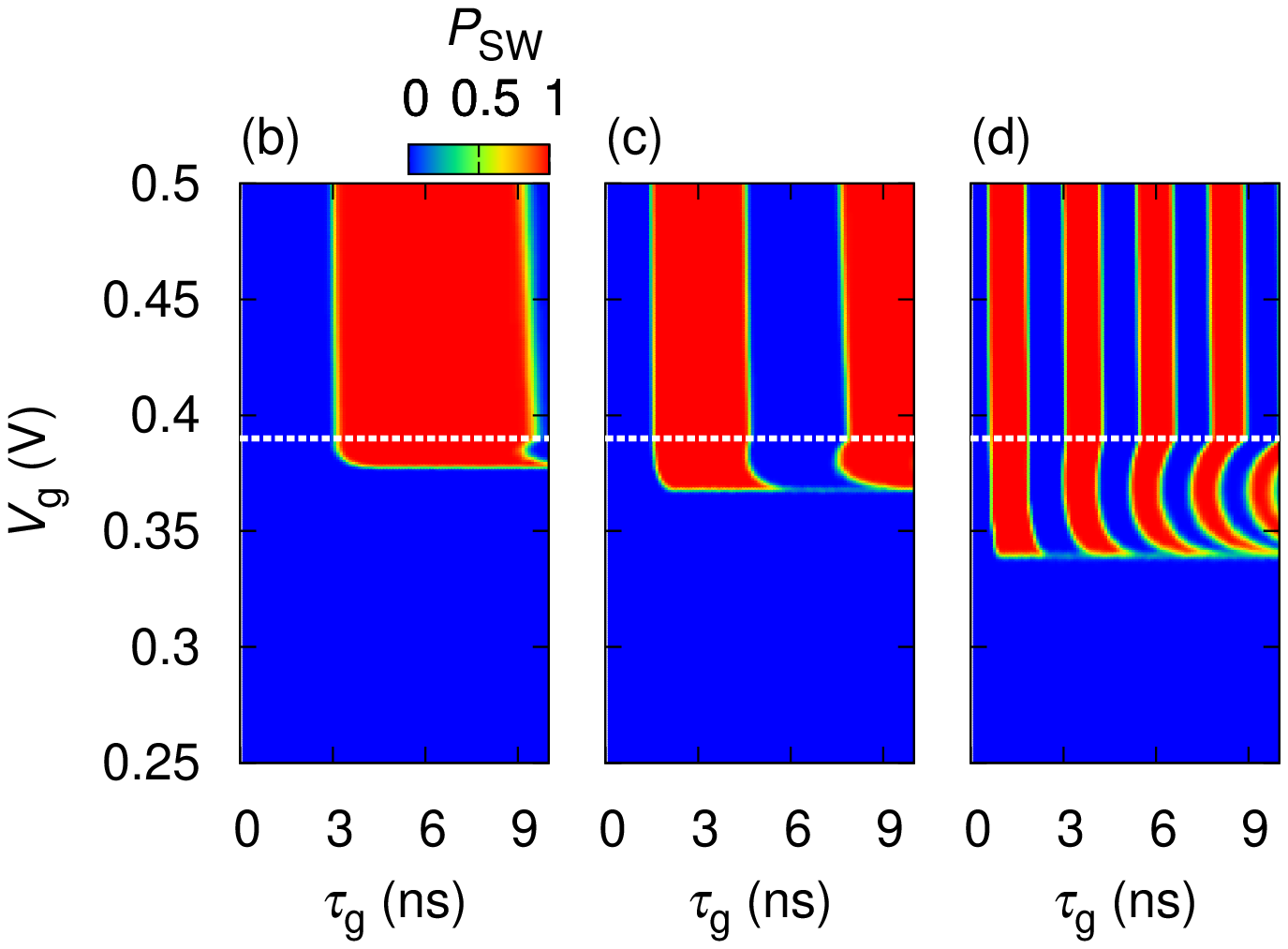}
 \end{center}
 \caption{(a) Switching probability $P_{\rm SW}$ as a function
 of source-drain electric field $E_{\rm sd}$ and gate pulse duration
 $\tau_{g}$ with ideal
 square pulse.
 The gate voltage is set to $0.39~{\rm V}$.
 The Gilbert damping $\alpha$ is set to 0.1 and the temperature
 $T$ is set to 300~{\rm K}.
 Magnetization switching probabilities as a
 function of gate voltage $V_{g}$ and gate pulse duration $\tau_{g}$
 with ideal square pulse. The temperature $T$ is set to 300~{\rm K}.
 Figures (b), (c), and (d) correspond to the parameter sets
 (1), (2), and (3) shown in (a), respectively.
 }
 \label{fig:diagram}
\end{figure}

Figure \ref{fig:diagram}(a) shows the switching probability $P_{\rm SW}$
as a function of gate pulse duration $\tau_{g}$ and source-drain
electric field $E_{\rm sd}$.
The switching probability $P_{\rm SW} = N_{\rm sw}/N_{\rm tot}$ 
was calculated from the number of successful switchings, $N_{\rm sw}$, 
against a million of gate voltage inputs, $N_{\rm tot}$,
without electrical noise under thermal fluctuation of ${\rm 300~K}$.
Clearly, the switching probability oscillates depending on both
$\tau_{g}$ and $E_{\rm sd}$.
Consequently, switching will be achieved in the wide pulse duration.
The magnetization switching occurs due to characteristic frequency
determined by SOT or $E_{\rm sd}$.
Thus, the adequate pulse duration $\tau_{g}$ is determined by SOT
strength, and, in any case, the pulse duration $\tau_{g}$ and the
source-drain electric field $E_{\rm sd}$ provide wide operation window.
The operation window can be estimated as follows:
\begin{equation}
 \tau_{g} = \frac{(2n+1)\pi(1-\alpha^2)}{\gamma B_{\rm SOT}},
  ~\left( n = 0,~1,~2,~\cdots \right) ,
  \label{eqn:period}
\end{equation}
which means the source-drain electric field can manipulate the operation
window or the desired pulse duration.
Figures \ref{fig:diagram}(b)-(d) also show the switching probability
$P_{\rm SW}$ as a function of gate pulse duration $\tau_{g}$ and
gate voltage $V_{g}$.
Clearly, the switching probability oscillates depending on only the
gate pulse width when the gate voltage is larger than 0.39~V
vanishing magnetic anisotropy.
The operation window period decreases as the SOT strength increases
according to Eq.(\ref{eqn:period}).
On the other hand, when the gate voltage is smaller than 0.39~V,
the distorted operation windows were observed due to the SOT.
The gate voltage which enables the magnetization switching becomes
smaller as the SOT strength increases.
These results are in good agreements with the estimated operation windows.
For parameter set (1) as shown in Fig.\ref{fig:diagram},
the source-drain electric field of $0.1~{\rm MV\,m^{-1}}$ is equivalent
to the $B_{\rm SOT}$ magnitude of about $12~{\rm mT}$.
The corresponding operation window $\tau_{g}$ is about $1.4~{\rm ns}$,
which is moderately high-speed and is easy to handle the write operation
by the present CMOS technology.
In order to investigate WERs,
three pairs of appropriate parameters of $\tau_{g}$ and $E_{\rm sd}$
were chosen as shown in Fig.\ref{fig:diagram}(a).
The WERs, $1 - N_{\rm sw}/N_{\rm tot}$, were also calculated from the
number of failure switching or write-error for various cut-off
frequencies without the other fluctuation.

\begin{figure*}[htp] 
 \begin{center}
  \includegraphics[width=4.8truecm,angle=270]{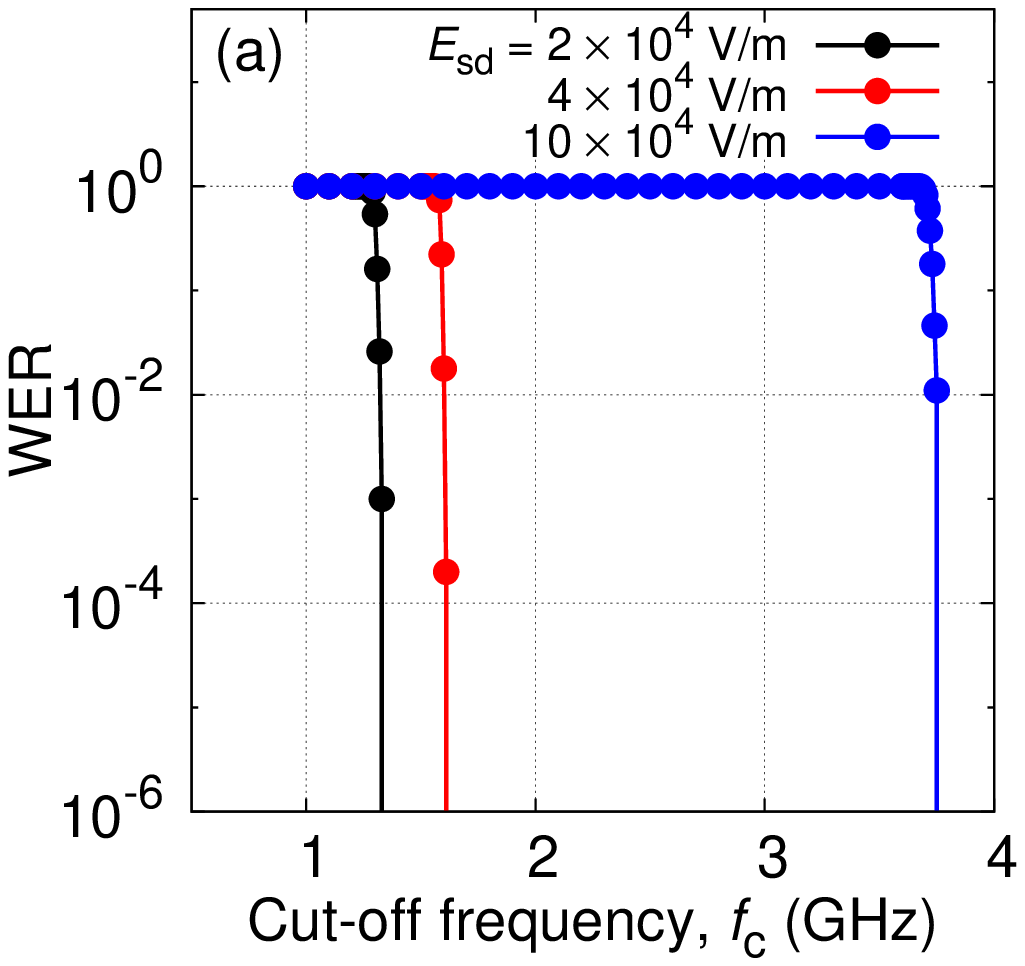}\ \ \ 
  \includegraphics[width=4.8truecm,angle=270]{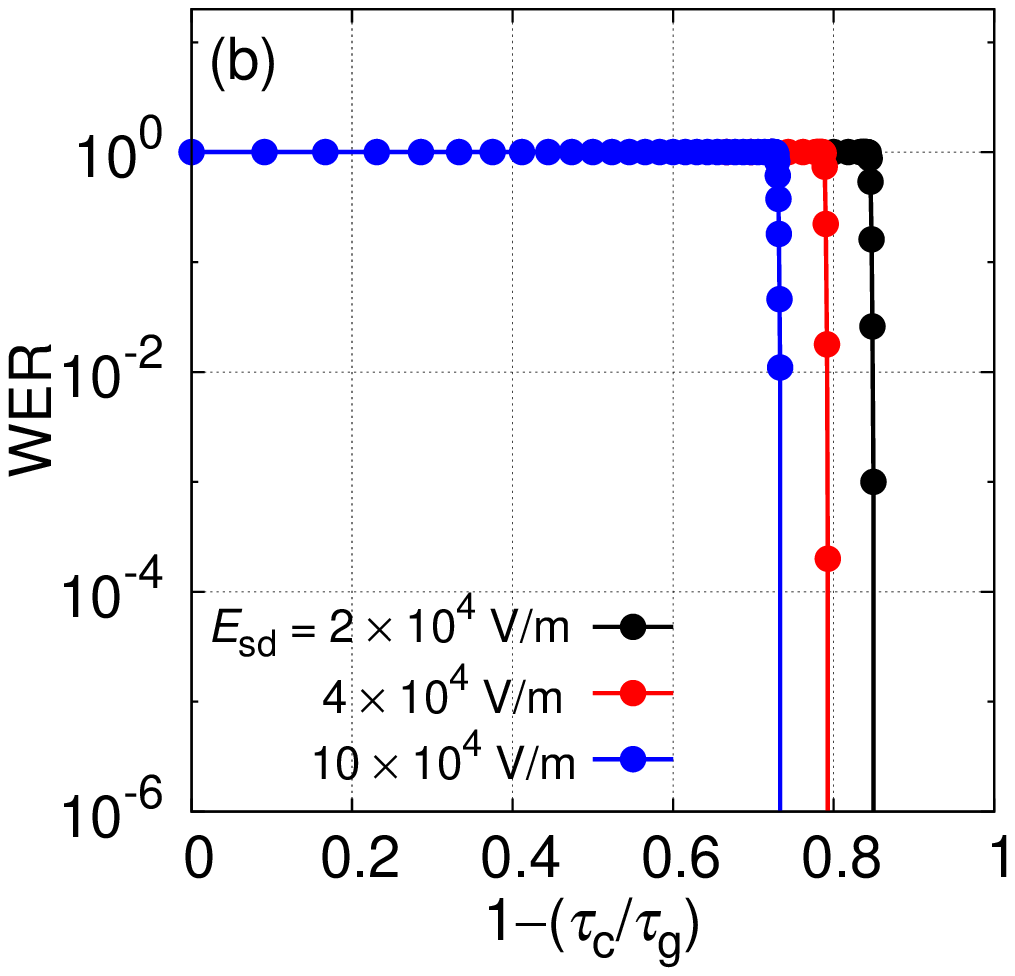}\ \ \ 
  \includegraphics[width=4.8truecm,angle=270]{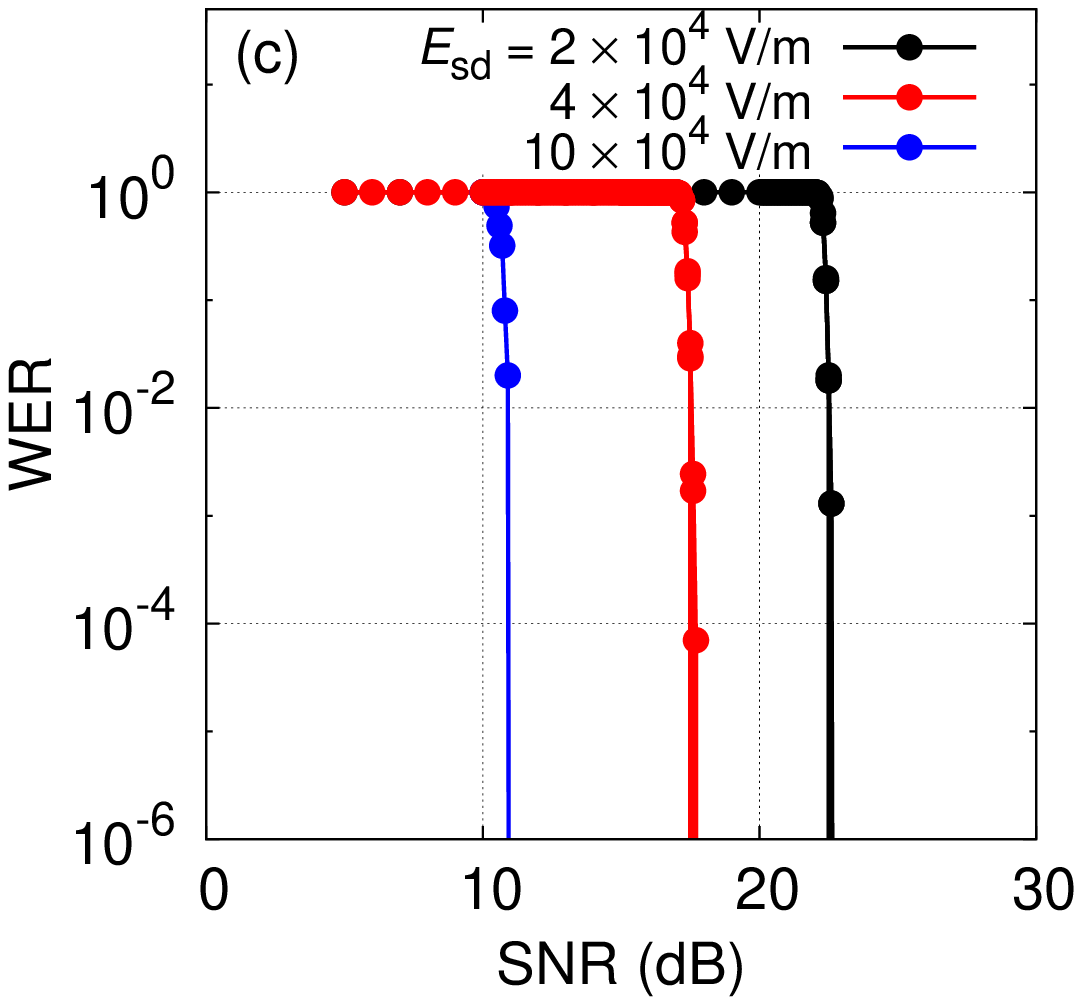}
 \end{center}
 \caption{(a) WER as a function of cut-off frequency $f_{c}$, and
 (b) WER as a function of $1-\tau_{c}/\tau_{g}$ which corresponds to the
 effective induction ratio of VCMA.
 (c) WER as a function of signal-to-noise ratio (SNR).}
 \label{fig:WER}
\end{figure*}
In order to investigate the electric circuit delay, the cut-off
frequency $f_{c}$ for gate pulse was varied and calculated WER.
Figure \ref{fig:WER} (a) shows cut-off frequency dependence of WER.
Since the adequate gate pulse $\tau_{g}$ becomes shorter as $E_{\rm sd}$
becomes larger, the required $f_{c}$ must be higher.
Since the rise time $\tau_{c}$ for 1st-order filter is defined to be
$1/f_{c}$, the ratio $1-\tau_{c}/\tau_{g}$ expresses the effective
induction ratio of VCMA as shown in Fig.\ref{fig:WER} (b).
If the rise time is about $15\%$ smaller than the gate pulse width,
the excellent WER can be obtained in each gate pulse.
The appropriate gate pulse can be chosen from the cut-off frequency in
the practical electric circuit.

\begin{table}
 \caption{Thermal stability estimated from calculated WERs as shown in
 Fig.\ref{fig:WER}(d).
 The device size $L$ of $100~{\rm nm}$ was used in these results.}
 \label{tbl:kuv}
 \centering
 \begin{tabular}{p{10em}p{10em}}
  \hline
  Source-drain field & Thermal stability \\
  $E_{\rm sd}~[{\rm MV\,m^{-1}}]$ & $K_{u}V/k_{\rm B}T~(T=300~{\rm K})$ \\
  \hline
  0.02 & 4.3 \\
  0.04 & 14.7 \\
  0.1 & 20.3 \\
  0.1,~$(\alpha=0.01)$ & 40.2 \\
  \hline
 \end{tabular}
\end{table}

\begin{figure}[htp] 
 \begin{center}
  \includegraphics[width=5.3truecm,angle=270]{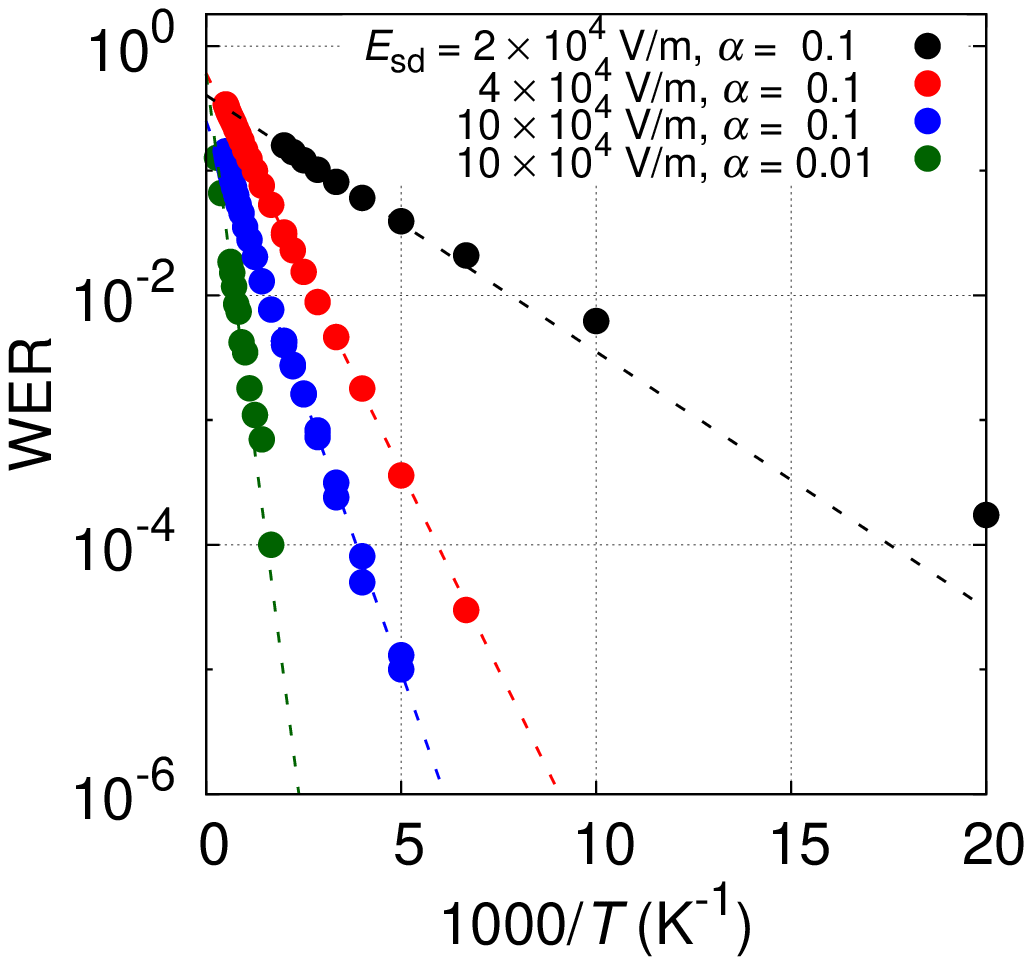}
 \end{center}
 \caption{Write-error-rate (WER) as a function of temperature $T$.
 The device size $L$ of $100~{\rm nm}$ was used.}
 \label{fig:WER-Temp}
\end{figure}

We also investigated the SNR dependence of WER for source-drain electric
field $E_{\rm sd}$.
The source-drain electric field $E_{\rm sd}$ is immune to noise, and the
required SNR is as very low as ${\rm -20dB}$.
Figure \ref{fig:WER} (c) shows WER as a function of SNR against the
gate voltage $V_{g}$. 
For larger $E_{\rm sd}$ which corresponds to long pulse duration
$\tau_{g}$, the required SNR of gate voltage becomes larger because of
the SOT assist by the source-drain electric field $E_{\rm sd}$. 
Both required cut-off frequency and SNRs can be sufficiently implemented
in the practical application of this device.
Thus, the MTI-based device is extremely robust against circuit delay and
signal-to-noise ratio.

Alike the voltage-controlled MTJ, the thermal fluctuation of
magnetization also gives rise an unexpected effect on the device
operation.
Since the thermal fluctuation effect on WER becomes more prominent as
the device size becomes smaller.
Thus, the device size is set to $100~{\rm nm}$ in the simulation.
Figure \ref{fig:WER-Temp} shows the temperature dependence of WER in
three different types of operation as shown in Fig.\ref{fig:diagram}.
Three type operations have different thermal stability factors.
As the larger SOT induces, the thermal stability factor improves more.
In the type (3) operation as shown in Fig.\ref{fig:diagram}, the gate
pulse duration is $1~{\rm ns}$ and high speed operation is a promising.
By fitting the slope of WER, the estimated thermal stability factors
$K_{u}V/k_{\rm B}T$ are shown in Table\ref{tbl:kuv}.
The estimated thermal stability factors
$K_{u}V/k_{\rm B}T$ is more than 40 in the device size of $100~{\rm
nm}$ and smaller damping constant $\alpha$, which are sufficient for the
practical application.
The SOT plays a role of the external magnetic field which is
similar to the external magnetic field in the VCMA-MTJ device.
The effective anisotropy field, $B^{*}_{k} = = \sqrt{{B_{k}}^2 + {B_{\rm
SOT}}^2}$, can be estimated, where $B_{k}$ and $B_{\rm SOT}$ are the
intrinsic anisotropy field and the field strength corresponding to the
SOT.
Thus, the thermal stability is enhanced during the application of SOT
as the SOT increases.

As the above-mentioned, the VCMA-MTI device has a feasible feature of
VCMA and superior WER characteristics comparing with VCMA-MTJ.
Furthermore, the writing energy of VCMA-MTI is less than that of
VCMA-MTJ due to steep change in VCMA of MTI.
On the other hand, the Curie temperature and thermal stability of MTI
are much less than that of materials in the MTJ device.
Since the thermal stability is related to the magnetically-opened
bandgap in MTI, development of potential materials is important
to realize the proposed MTI-based device.
Finally, we briefly mention about potential materials to fabricate the
MTI-based device. To realize MTIs, most of the past research has been
dedicated to Cr-doped TI, which restricts the Curie temperature to be
around 10-20 K\cite{Li_PhysLettA_2013,Chang_AdvMater_2013}.
For practical device applications, room-temperature MTIs would be 
essential and resent intensive studies for MTIs with other
magnetic dopants will facilitate the realization of the presented
device.

In summary, we theoretically investigate influences of electronic
circuit delay, noise and temperature on write-error-rate (WER) in
voltage-controlled magnetization switching operation of a MTI-based
device by means of the micromagnetic simulation. 
This device realizes magnetization switching via SOT and VCMA which
originate from 2D-Dirac electronic structure. 
We reveal that the device operation is extremely robust against circuit
delay and SNR.
We demonstrate that the WER on the order of approximately $10^{-4}$ or
below is achieved around room temperature due to steep change in VCMA. 
Also, we show that the larger SOT improves thermal stability factor. 
Therefore, this study provides a next perspective for developing 
voltage-driven spintronic devices with ultra-low power consumption.

\section*{Supplementary material}

See the supplementary material for the energy surface, trajectory,
and fully micromagnetic understanding during magnetization switching.

\vspace{1em}

This work was partly supported by Grants-in-Aid for Scientific Research
(Grants No. 20H02196, 22K14591, 22H01805, 20K03814, 18KK0132) from the
Japan Society for the Promotion of Science,
by the Spintronics Research Network of Japan (Spin-RNJ). 
This work was partially performed under the Research Program of
``Dynamic Alliance for Open Innovation Bridging Human, Environment and
Materials'' in ``Network Joint Research Center for Materials and
Devices''.

\bibliographystyle{apsrev4-1}
\bibliography{refs}

\begin{thebibliography}{45}%
\makeatletter
\providecommand \@ifxundefined [1]{%
 \@ifx{#1\undefined}
}%
\providecommand \@ifnum [1]{%
 \ifnum #1\expandafter \@firstoftwo
 \else \expandafter \@secondoftwo
 \fi
}%
\providecommand \@ifx [1]{%
 \ifx #1\expandafter \@firstoftwo
 \else \expandafter \@secondoftwo
 \fi
}%
\providecommand \natexlab [1]{#1}%
\providecommand \enquote  [1]{``#1''}%
\providecommand \bibnamefont  [1]{#1}%
\providecommand \bibfnamefont [1]{#1}%
\providecommand \citenamefont [1]{#1}%
\providecommand \href@noop [0]{\@secondoftwo}%
\providecommand \href [0]{\begingroup \@sanitize@url \@href}%
\providecommand \@href[1]{\@@startlink{#1}\@@href}%
\providecommand \@@href[1]{\endgroup#1\@@endlink}%
\providecommand \@sanitize@url [0]{\catcode `\\12\catcode `\$12\catcode
  `\&12\catcode `\#12\catcode `\^12\catcode `\_12\catcode `\%12\relax}%
\providecommand \@@startlink[1]{}%
\providecommand \@@endlink[0]{}%
\providecommand \url  [0]{\begingroup\@sanitize@url \@url }%
\providecommand \@url [1]{\endgroup\@href {#1}{\urlprefix }}%
\providecommand \urlprefix  [0]{URL }%
\providecommand \Eprint [0]{\href }%
\providecommand \doibase [0]{http://dx.doi.org/}%
\providecommand \selectlanguage [0]{\@gobble}%
\providecommand \bibinfo  [0]{\@secondoftwo}%
\providecommand \bibfield  [0]{\@secondoftwo}%
\providecommand \translation [1]{[#1]}%
\providecommand \BibitemOpen [0]{}%
\providecommand \bibitemStop [0]{}%
\providecommand \bibitemNoStop [0]{.\EOS\space}%
\providecommand \EOS [0]{\spacefactor3000\relax}%
\providecommand \BibitemShut  [1]{\csname bibitem#1\endcsname}%
\let\auto@bib@innerbib\@empty
\bibitem [{\citenamefont {Ando}\ \emph {et~al.}(2014)\citenamefont {Ando},
  \citenamefont {Fujita}, \citenamefont {Ito}, \citenamefont {Yuasa},
  \citenamefont {Suzuki}, \citenamefont {Nakatani}, \citenamefont {Miyazaki},\
  and\ \citenamefont {Yoda}}]{Ando_JAP_2014}%
  \BibitemOpen
  \bibfield  {author} {\bibinfo {author} {\bibfnamefont {K.}~\bibnamefont
  {Ando}}, \bibinfo {author} {\bibfnamefont {S.}~\bibnamefont {Fujita}},
  \bibinfo {author} {\bibfnamefont {J.}~\bibnamefont {Ito}}, \bibinfo {author}
  {\bibfnamefont {S.}~\bibnamefont {Yuasa}}, \bibinfo {author} {\bibfnamefont
  {Y.}~\bibnamefont {Suzuki}}, \bibinfo {author} {\bibfnamefont
  {Y.}~\bibnamefont {Nakatani}}, \bibinfo {author} {\bibfnamefont
  {T.}~\bibnamefont {Miyazaki}}, \ and\ \bibinfo {author} {\bibfnamefont
  {H.}~\bibnamefont {Yoda}},\ }\href {\doibase 10.1063/1.4869828} {\bibfield
  {journal} {\bibinfo  {journal} {Journal of Applied Physics}\ }\textbf
  {\bibinfo {volume} {115}} (\bibinfo {year} {2014}),\
  10.1063/1.4869828}\BibitemShut {NoStop}%
\bibitem [{\citenamefont {Barla}\ \emph {et~al.}(2021)\citenamefont {Barla},
  \citenamefont {Joshi},\ and\ \citenamefont {Bhat}}]{Barla_JCompElec_2021}%
  \BibitemOpen
  \bibfield  {author} {\bibinfo {author} {\bibfnamefont {P.}~\bibnamefont
  {Barla}}, \bibinfo {author} {\bibfnamefont {V.~K.}\ \bibnamefont {Joshi}}, \
  and\ \bibinfo {author} {\bibfnamefont {S.}~\bibnamefont {Bhat}},\ }\href
  {\doibase 10.1007/s10825-020-01648-6} {\bibfield  {journal} {\bibinfo
  {journal} {J. Comp. Electron.}\ }\textbf {\bibinfo {volume} {20}},\ \bibinfo
  {pages} {805} (\bibinfo {year} {2021})}\BibitemShut {NoStop}%
\bibitem [{\citenamefont {Zhang}\ \emph {et~al.}(2014)\citenamefont {Zhang},
  \citenamefont {Zhao}, \citenamefont {Klein}, \citenamefont {Kang},
  \citenamefont {Querlioz}, \citenamefont {Zhang}, \citenamefont {Ravelosona},\
  and\ \citenamefont {Chappert}}]{Zhang_IEEE_6800517}%
  \BibitemOpen
  \bibfield  {author} {\bibinfo {author} {\bibfnamefont {Y.}~\bibnamefont
  {Zhang}}, \bibinfo {author} {\bibfnamefont {W.}~\bibnamefont {Zhao}},
  \bibinfo {author} {\bibfnamefont {J.-O.}\ \bibnamefont {Klein}}, \bibinfo
  {author} {\bibfnamefont {W.}~\bibnamefont {Kang}}, \bibinfo {author}
  {\bibfnamefont {D.}~\bibnamefont {Querlioz}}, \bibinfo {author}
  {\bibfnamefont {Y.}~\bibnamefont {Zhang}}, \bibinfo {author} {\bibfnamefont
  {D.}~\bibnamefont {Ravelosona}}, \ and\ \bibinfo {author} {\bibfnamefont
  {C.}~\bibnamefont {Chappert}},\ }in\ \href {\doibase 10.7873/DATE.2014.316}
  {\emph {\bibinfo {booktitle} {2014 Design, Automation \& Test in Europe
  Conference \& Exhibition (DATE)}}}\ (\bibinfo {year} {2014})\ pp.\ \bibinfo
  {pages} {1--6}\BibitemShut {NoStop}%
\bibitem [{\citenamefont {Maruyama}\ \emph {et~al.}(2009)\citenamefont
  {Maruyama}, \citenamefont {Shiota}, \citenamefont {Nozaki}, \citenamefont
  {Ohta}, \citenamefont {Toda}, \citenamefont {Mizuguchi}, \citenamefont
  {Tulapurkar}, \citenamefont {Shinjo}, \citenamefont {Shiraishi},
  \citenamefont {Mizukami}, \citenamefont {Ando},\ and\ \citenamefont
  {Suzuki}}]{Maruyama_NatNano_2009}%
  \BibitemOpen
  \bibfield  {author} {\bibinfo {author} {\bibfnamefont {T.}~\bibnamefont
  {Maruyama}}, \bibinfo {author} {\bibfnamefont {Y.}~\bibnamefont {Shiota}},
  \bibinfo {author} {\bibfnamefont {T.}~\bibnamefont {Nozaki}}, \bibinfo
  {author} {\bibfnamefont {K.}~\bibnamefont {Ohta}}, \bibinfo {author}
  {\bibfnamefont {N.}~\bibnamefont {Toda}}, \bibinfo {author} {\bibfnamefont
  {M.}~\bibnamefont {Mizuguchi}}, \bibinfo {author} {\bibfnamefont {A.~A.}\
  \bibnamefont {Tulapurkar}}, \bibinfo {author} {\bibfnamefont
  {T.}~\bibnamefont {Shinjo}}, \bibinfo {author} {\bibfnamefont
  {M.}~\bibnamefont {Shiraishi}}, \bibinfo {author} {\bibfnamefont
  {S.}~\bibnamefont {Mizukami}}, \bibinfo {author} {\bibfnamefont
  {Y.}~\bibnamefont {Ando}}, \ and\ \bibinfo {author} {\bibfnamefont
  {Y.}~\bibnamefont {Suzuki}},\ }\href {\doibase 10.1038/nnano.2008.406}
  {\bibfield  {journal} {\bibinfo  {journal} {Nat. Nanotech.}\ }\textbf
  {\bibinfo {volume} {4}},\ \bibinfo {pages} {158–161} (\bibinfo {year}
  {2009})}\BibitemShut {NoStop}%
\bibitem [{\citenamefont {Endo}\ \emph {et~al.}(2010)\citenamefont {Endo},
  \citenamefont {Kanai}, \citenamefont {Ikeda}, \citenamefont {Matsukura},\
  and\ \citenamefont {Ohno}}]{Endo_APL_2010}%
  \BibitemOpen
  \bibfield  {author} {\bibinfo {author} {\bibfnamefont {M.}~\bibnamefont
  {Endo}}, \bibinfo {author} {\bibfnamefont {S.}~\bibnamefont {Kanai}},
  \bibinfo {author} {\bibfnamefont {S.}~\bibnamefont {Ikeda}}, \bibinfo
  {author} {\bibfnamefont {F.}~\bibnamefont {Matsukura}}, \ and\ \bibinfo
  {author} {\bibfnamefont {H.}~\bibnamefont {Ohno}},\ }\href {\doibase
  10.1063/1.3429592} {\bibfield  {journal} {\bibinfo  {journal} {Appl. Phys.
  Lett.}\ }\textbf {\bibinfo {volume} {96}},\ \bibinfo {pages} {212503}
  (\bibinfo {year} {2010})}\BibitemShut {NoStop}%
\bibitem [{\citenamefont {Shiota}\ \emph {et~al.}(2011)\citenamefont {Shiota},
  \citenamefont {Nozaki}, \citenamefont {Bonell}, \citenamefont {Murakami},
  \citenamefont {Shinjo},\ and\ \citenamefont {Suzuki}}]{Shiota_NatMater_2011}%
  \BibitemOpen
  \bibfield  {author} {\bibinfo {author} {\bibfnamefont {Y.}~\bibnamefont
  {Shiota}}, \bibinfo {author} {\bibfnamefont {T.}~\bibnamefont {Nozaki}},
  \bibinfo {author} {\bibfnamefont {F.}~\bibnamefont {Bonell}}, \bibinfo
  {author} {\bibfnamefont {S.}~\bibnamefont {Murakami}}, \bibinfo {author}
  {\bibfnamefont {T.}~\bibnamefont {Shinjo}}, \ and\ \bibinfo {author}
  {\bibfnamefont {Y.}~\bibnamefont {Suzuki}},\ }\href {\doibase
  10.1038/nmat3172} {\bibfield  {journal} {\bibinfo  {journal} {Nat. Mater.}\
  }\textbf {\bibinfo {volume} {11}},\ \bibinfo {pages} {39} (\bibinfo {year}
  {2011})}\BibitemShut {NoStop}%
\bibitem [{\citenamefont {Shiota}\ \emph {et~al.}(2012)\citenamefont {Shiota},
  \citenamefont {Miwa}, \citenamefont {Nozaki}, \citenamefont {Bonell},
  \citenamefont {Mizuochi}, \citenamefont {Shinjo}, \citenamefont {Kubota},
  \citenamefont {Yuasa},\ and\ \citenamefont {Suzuki}}]{Shiota_APL_2012}%
  \BibitemOpen
  \bibfield  {author} {\bibinfo {author} {\bibfnamefont {Y.}~\bibnamefont
  {Shiota}}, \bibinfo {author} {\bibfnamefont {S.}~\bibnamefont {Miwa}},
  \bibinfo {author} {\bibfnamefont {T.}~\bibnamefont {Nozaki}}, \bibinfo
  {author} {\bibfnamefont {F.}~\bibnamefont {Bonell}}, \bibinfo {author}
  {\bibfnamefont {N.}~\bibnamefont {Mizuochi}}, \bibinfo {author}
  {\bibfnamefont {T.}~\bibnamefont {Shinjo}}, \bibinfo {author} {\bibfnamefont
  {H.}~\bibnamefont {Kubota}}, \bibinfo {author} {\bibfnamefont
  {S.}~\bibnamefont {Yuasa}}, \ and\ \bibinfo {author} {\bibfnamefont
  {Y.}~\bibnamefont {Suzuki}},\ }\href {\doibase 10.1063/1.4751035} {\bibfield
  {journal} {\bibinfo  {journal} {Appl. Phys. Lett.}\ }\textbf {\bibinfo
  {volume} {101}},\ \bibinfo {pages} {102406} (\bibinfo {year}
  {2012})}\BibitemShut {NoStop}%
\bibitem [{\citenamefont {Shiota}\ \emph {et~al.}(2015)\citenamefont {Shiota},
  \citenamefont {Nozaki}, \citenamefont {Tamaru}, \citenamefont {Yakushiji},
  \citenamefont {Kubota}, \citenamefont {Fukushima}, \citenamefont {Yuasa},\
  and\ \citenamefont {Suzuki}}]{Shiota_APEX_2016}%
  \BibitemOpen
  \bibfield  {author} {\bibinfo {author} {\bibfnamefont {Y.}~\bibnamefont
  {Shiota}}, \bibinfo {author} {\bibfnamefont {T.}~\bibnamefont {Nozaki}},
  \bibinfo {author} {\bibfnamefont {S.}~\bibnamefont {Tamaru}}, \bibinfo
  {author} {\bibfnamefont {K.}~\bibnamefont {Yakushiji}}, \bibinfo {author}
  {\bibfnamefont {H.}~\bibnamefont {Kubota}}, \bibinfo {author} {\bibfnamefont
  {A.}~\bibnamefont {Fukushima}}, \bibinfo {author} {\bibfnamefont
  {S.}~\bibnamefont {Yuasa}}, \ and\ \bibinfo {author} {\bibfnamefont
  {Y.}~\bibnamefont {Suzuki}},\ }\href {\doibase 10.7567/APEX.9.013001}
  {\bibfield  {journal} {\bibinfo  {journal} {Appl. Phys. Express}\ }\textbf
  {\bibinfo {volume} {9}},\ \bibinfo {pages} {013001} (\bibinfo {year}
  {2015})}\BibitemShut {NoStop}%
\bibitem [{\citenamefont {Grezes}\ \emph {et~al.}(2016)\citenamefont {Grezes},
  \citenamefont {Ebrahimi}, \citenamefont {Alzate}, \citenamefont {Cai},
  \citenamefont {Katine}, \citenamefont {Langer}, \citenamefont {Ocker},
  \citenamefont {Khalili~Amiri},\ and\ \citenamefont {Wang}}]{Grezes_APL_2016}%
  \BibitemOpen
  \bibfield  {author} {\bibinfo {author} {\bibfnamefont {C.}~\bibnamefont
  {Grezes}}, \bibinfo {author} {\bibfnamefont {F.}~\bibnamefont {Ebrahimi}},
  \bibinfo {author} {\bibfnamefont {J.~G.}\ \bibnamefont {Alzate}}, \bibinfo
  {author} {\bibfnamefont {X.}~\bibnamefont {Cai}}, \bibinfo {author}
  {\bibfnamefont {J.~A.}\ \bibnamefont {Katine}}, \bibinfo {author}
  {\bibfnamefont {J.}~\bibnamefont {Langer}}, \bibinfo {author} {\bibfnamefont
  {B.}~\bibnamefont {Ocker}}, \bibinfo {author} {\bibfnamefont
  {P.}~\bibnamefont {Khalili~Amiri}}, \ and\ \bibinfo {author} {\bibfnamefont
  {K.~L.}\ \bibnamefont {Wang}},\ }\href {\doibase 10.1063/1.4939446}
  {\bibfield  {journal} {\bibinfo  {journal} {Appl. Phys. Lett.}\ }\textbf
  {\bibinfo {volume} {108}},\ \bibinfo {pages} {012403} (\bibinfo {year}
  {2016})}\BibitemShut {NoStop}%
\bibitem [{\citenamefont {Yamamoto}\ \emph {et~al.}(2022)\citenamefont
  {Yamamoto}, \citenamefont {Matsumoto}, \citenamefont {Nozaki}, \citenamefont
  {Imamura},\ and\ \citenamefont {Yuasa}}]{VC-MTJ-review}%
  \BibitemOpen
  \bibfield  {author} {\bibinfo {author} {\bibfnamefont {T.}~\bibnamefont
  {Yamamoto}}, \bibinfo {author} {\bibfnamefont {R.}~\bibnamefont {Matsumoto}},
  \bibinfo {author} {\bibfnamefont {T.}~\bibnamefont {Nozaki}}, \bibinfo
  {author} {\bibfnamefont {H.}~\bibnamefont {Imamura}}, \ and\ \bibinfo
  {author} {\bibfnamefont {S.}~\bibnamefont {Yuasa}},\ }\href@noop {}
  {\bibfield  {journal} {\bibinfo  {journal} {J. Magn. Magn. Mater.}\ }\textbf
  {\bibinfo {volume} {560}},\ \bibinfo {pages} {169637} (\bibinfo {year}
  {2022})}\BibitemShut {NoStop}%
\bibitem [{\citenamefont {Shiota}\ \emph {et~al.}(2017)\citenamefont {Shiota},
  \citenamefont {Nozaki}, \citenamefont {Tamaru}, \citenamefont {Yakushiji},
  \citenamefont {Kubota}, \citenamefont {Fukushima}, \citenamefont {Yuasa},\
  and\ \citenamefont {Suzuki}}]{Shiota_APL_2017}%
  \BibitemOpen
  \bibfield  {author} {\bibinfo {author} {\bibfnamefont {Y.}~\bibnamefont
  {Shiota}}, \bibinfo {author} {\bibfnamefont {T.}~\bibnamefont {Nozaki}},
  \bibinfo {author} {\bibfnamefont {S.}~\bibnamefont {Tamaru}}, \bibinfo
  {author} {\bibfnamefont {K.}~\bibnamefont {Yakushiji}}, \bibinfo {author}
  {\bibfnamefont {H.}~\bibnamefont {Kubota}}, \bibinfo {author} {\bibfnamefont
  {A.}~\bibnamefont {Fukushima}}, \bibinfo {author} {\bibfnamefont
  {S.}~\bibnamefont {Yuasa}}, \ and\ \bibinfo {author} {\bibfnamefont
  {Y.}~\bibnamefont {Suzuki}},\ }\href {\doibase 10.1063/1.4990680} {\bibfield
  {journal} {\bibinfo  {journal} {Appl. Phys. Lett.}\ }\textbf {\bibinfo
  {volume} {111}},\ \bibinfo {pages} {022408} (\bibinfo {year}
  {2017})}\BibitemShut {NoStop}%
\bibitem [{\citenamefont {Yamamoto}\ \emph {et~al.}(2019)\citenamefont
  {Yamamoto}, \citenamefont {Nozaki}, \citenamefont {Imamura}, \citenamefont
  {Shiota}, \citenamefont {Ikeura}, \citenamefont {Tamaru}, \citenamefont
  {Yakushiji}, \citenamefont {Kubota}, \citenamefont {Fukushima}, \citenamefont
  {Suzuki},\ and\ \citenamefont {Yuasa}}]{Yamamoto_PRA_2019}%
  \BibitemOpen
  \bibfield  {author} {\bibinfo {author} {\bibfnamefont {T.}~\bibnamefont
  {Yamamoto}}, \bibinfo {author} {\bibfnamefont {T.}~\bibnamefont {Nozaki}},
  \bibinfo {author} {\bibfnamefont {H.}~\bibnamefont {Imamura}}, \bibinfo
  {author} {\bibfnamefont {Y.}~\bibnamefont {Shiota}}, \bibinfo {author}
  {\bibfnamefont {T.}~\bibnamefont {Ikeura}}, \bibinfo {author} {\bibfnamefont
  {S.}~\bibnamefont {Tamaru}}, \bibinfo {author} {\bibfnamefont
  {K.}~\bibnamefont {Yakushiji}}, \bibinfo {author} {\bibfnamefont
  {H.}~\bibnamefont {Kubota}}, \bibinfo {author} {\bibfnamefont
  {A.}~\bibnamefont {Fukushima}}, \bibinfo {author} {\bibfnamefont
  {Y.}~\bibnamefont {Suzuki}}, \ and\ \bibinfo {author} {\bibfnamefont
  {S.}~\bibnamefont {Yuasa}},\ }\href {\doibase
  10.1103/PhysRevApplied.11.014013} {\bibfield  {journal} {\bibinfo  {journal}
  {Phys. Rev. Appl.}\ }\textbf {\bibinfo {volume} {11}},\ \bibinfo {pages}
  {014013} (\bibinfo {year} {2019})}\BibitemShut {NoStop}%
\bibitem [{\citenamefont {Ando}(2013)}]{Ando_2013}%
  \BibitemOpen
  \bibfield  {author} {\bibinfo {author} {\bibfnamefont {Y.}~\bibnamefont
  {Ando}},\ }\href {\doibase 10.7566/JPSJ.82.102001} {\bibfield  {journal}
  {\bibinfo  {journal} {J. Phys. Soc. Jpn.}\ }\textbf {\bibinfo {volume}
  {82}},\ \bibinfo {pages} {102001} (\bibinfo {year} {2013})}\BibitemShut
  {NoStop}%
\bibitem [{\citenamefont {Tokura}\ \emph {et~al.}(2019)\citenamefont {Tokura},
  \citenamefont {Yasuda},\ and\ \citenamefont {Tsukazaki}}]{Tokura_2019}%
  \BibitemOpen
  \bibfield  {author} {\bibinfo {author} {\bibfnamefont {Y.}~\bibnamefont
  {Tokura}}, \bibinfo {author} {\bibfnamefont {K.}~\bibnamefont {Yasuda}}, \
  and\ \bibinfo {author} {\bibfnamefont {A.}~\bibnamefont {Tsukazaki}},\ }\href
  {\doibase 10.1038/s42254-018-0011-5} {\bibfield  {journal} {\bibinfo
  {journal} {Nat. Rev. Phys.}\ }\textbf {\bibinfo {volume} {1}},\ \bibinfo
  {pages} {126} (\bibinfo {year} {2019})}\BibitemShut {NoStop}%
\bibitem [{\citenamefont {Wang}\ \emph {et~al.}(2015)\citenamefont {Wang},
  \citenamefont {Lian},\ and\ \citenamefont {Zhang}}]{Wang_PRL_2015}%
  \BibitemOpen
  \bibfield  {author} {\bibinfo {author} {\bibfnamefont {J.}~\bibnamefont
  {Wang}}, \bibinfo {author} {\bibfnamefont {B.}~\bibnamefont {Lian}}, \ and\
  \bibinfo {author} {\bibfnamefont {S.-C.}\ \bibnamefont {Zhang}},\ }\href
  {\doibase 10.1103/PhysRevLett.115.036805} {\bibfield  {journal} {\bibinfo
  {journal} {Phys. Rev. Lett.}\ }\textbf {\bibinfo {volume} {115}},\ \bibinfo
  {pages} {036805} (\bibinfo {year} {2015})}\BibitemShut {NoStop}%
\bibitem [{\citenamefont {Sekine}\ and\ \citenamefont
  {Chiba}(2016)}]{Sekine_PRB_2016}%
  \BibitemOpen
  \bibfield  {author} {\bibinfo {author} {\bibfnamefont {A.}~\bibnamefont
  {Sekine}}\ and\ \bibinfo {author} {\bibfnamefont {T.}~\bibnamefont {Chiba}},\
  }\href {\doibase 10.1103/PhysRevB.93.220403} {\bibfield  {journal} {\bibinfo
  {journal} {Phys. Rev. B}\ }\textbf {\bibinfo {volume} {93}},\ \bibinfo
  {pages} {220403} (\bibinfo {year} {2016})}\BibitemShut {NoStop}%
\bibitem [{\citenamefont {Semenov}\ \emph {et~al.}(2012)\citenamefont
  {Semenov}, \citenamefont {Duan},\ and\ \citenamefont
  {Kim}}]{Semenov_PRB_2012}%
  \BibitemOpen
  \bibfield  {author} {\bibinfo {author} {\bibfnamefont {Y.~G.}\ \bibnamefont
  {Semenov}}, \bibinfo {author} {\bibfnamefont {X.}~\bibnamefont {Duan}}, \
  and\ \bibinfo {author} {\bibfnamefont {K.~W.}\ \bibnamefont {Kim}},\ }\href
  {\doibase 10.1103/PhysRevB.86.161406} {\bibfield  {journal} {\bibinfo
  {journal} {Phys. Rev. B}\ }\textbf {\bibinfo {volume} {86}},\ \bibinfo
  {pages} {161406} (\bibinfo {year} {2012})}\BibitemShut {NoStop}%
\bibitem [{\citenamefont {Fan}\ \emph {et~al.}(2016{\natexlab{a}})\citenamefont
  {Fan}, \citenamefont {Kou}, \citenamefont {Upadhyaya}, \citenamefont {Shao},
  \citenamefont {Pan}, \citenamefont {Lang}, \citenamefont {Che}, \citenamefont
  {Tang}, \citenamefont {Montazeri}, \citenamefont {Murata}, \citenamefont
  {Chang}, \citenamefont {Akyol}, \citenamefont {Yu}, \citenamefont {Nie},
  \citenamefont {Wong}, \citenamefont {Liu}, \citenamefont {Wang},
  \citenamefont {Tserkovnyak},\ and\ \citenamefont {Wang}}]{Fan_2016}%
  \BibitemOpen
  \bibfield  {author} {\bibinfo {author} {\bibfnamefont {Y.}~\bibnamefont
  {Fan}}, \bibinfo {author} {\bibfnamefont {X.}~\bibnamefont {Kou}}, \bibinfo
  {author} {\bibfnamefont {P.}~\bibnamefont {Upadhyaya}}, \bibinfo {author}
  {\bibfnamefont {Q.}~\bibnamefont {Shao}}, \bibinfo {author} {\bibfnamefont
  {L.}~\bibnamefont {Pan}}, \bibinfo {author} {\bibfnamefont {M.}~\bibnamefont
  {Lang}}, \bibinfo {author} {\bibfnamefont {X.}~\bibnamefont {Che}}, \bibinfo
  {author} {\bibfnamefont {J.}~\bibnamefont {Tang}}, \bibinfo {author}
  {\bibfnamefont {M.}~\bibnamefont {Montazeri}}, \bibinfo {author}
  {\bibfnamefont {K.}~\bibnamefont {Murata}}, \bibinfo {author} {\bibfnamefont
  {L.-T.}\ \bibnamefont {Chang}}, \bibinfo {author} {\bibfnamefont
  {M.}~\bibnamefont {Akyol}}, \bibinfo {author} {\bibfnamefont
  {G.}~\bibnamefont {Yu}}, \bibinfo {author} {\bibfnamefont {T.}~\bibnamefont
  {Nie}}, \bibinfo {author} {\bibfnamefont {K.~L.}\ \bibnamefont {Wong}},
  \bibinfo {author} {\bibfnamefont {J.}~\bibnamefont {Liu}}, \bibinfo {author}
  {\bibfnamefont {Y.}~\bibnamefont {Wang}}, \bibinfo {author} {\bibfnamefont
  {Y.}~\bibnamefont {Tserkovnyak}}, \ and\ \bibinfo {author} {\bibfnamefont
  {K.~L.}\ \bibnamefont {Wang}},\ }\href {\doibase 10.1038/nnano.2015.294}
  {\bibfield  {journal} {\bibinfo  {journal} {Nat. Nanotech.}\ }\textbf
  {\bibinfo {volume} {11}},\ \bibinfo {pages} {352} (\bibinfo {year}
  {2016}{\natexlab{a}})}\BibitemShut {NoStop}%
\bibitem [{\citenamefont {Khang}\ \emph {et~al.}(2018)\citenamefont {Khang},
  \citenamefont {Ueda},\ and\ \citenamefont {Hai}}]{Pham_Nature_2018}%
  \BibitemOpen
  \bibfield  {author} {\bibinfo {author} {\bibfnamefont {N.~H.~D.}\
  \bibnamefont {Khang}}, \bibinfo {author} {\bibfnamefont {Y.}~\bibnamefont
  {Ueda}}, \ and\ \bibinfo {author} {\bibfnamefont {P.~N.}\ \bibnamefont
  {Hai}},\ }\href {\doibase 10.1038/s41563-018-0137-y} {\bibfield  {journal}
  {\bibinfo  {journal} {Nat. Mater.}\ }\textbf {\bibinfo {volume} {17}},\
  \bibinfo {pages} {808} (\bibinfo {year} {2018})}\BibitemShut {NoStop}%
\bibitem [{\citenamefont {Cai}\ \emph {et~al.}(2017)\citenamefont {Cai},
  \citenamefont {Yang}, \citenamefont {Ju}, \citenamefont {Wang}, \citenamefont
  {Ji}, \citenamefont {Li}, \citenamefont {Edmonds}, \citenamefont {Sheng},
  \citenamefont {Zhang}, \citenamefont {Zhang}, \citenamefont {Liu},
  \citenamefont {Zheng},\ and\ \citenamefont {Wang}}]{nmatCai2017}%
  \BibitemOpen
  \bibfield  {author} {\bibinfo {author} {\bibfnamefont {K.}~\bibnamefont
  {Cai}}, \bibinfo {author} {\bibfnamefont {M.}~\bibnamefont {Yang}}, \bibinfo
  {author} {\bibfnamefont {H.}~\bibnamefont {Ju}}, \bibinfo {author}
  {\bibfnamefont {S.}~\bibnamefont {Wang}}, \bibinfo {author} {\bibfnamefont
  {Y.}~\bibnamefont {Ji}}, \bibinfo {author} {\bibfnamefont {B.}~\bibnamefont
  {Li}}, \bibinfo {author} {\bibfnamefont {K.~W.}\ \bibnamefont {Edmonds}},
  \bibinfo {author} {\bibfnamefont {Y.}~\bibnamefont {Sheng}}, \bibinfo
  {author} {\bibfnamefont {B.}~\bibnamefont {Zhang}}, \bibinfo {author}
  {\bibfnamefont {N.}~\bibnamefont {Zhang}}, \bibinfo {author} {\bibfnamefont
  {S.}~\bibnamefont {Liu}}, \bibinfo {author} {\bibfnamefont {H.}~\bibnamefont
  {Zheng}}, \ and\ \bibinfo {author} {\bibfnamefont {K.}~\bibnamefont {Wang}},\
  }\href {\doibase 10.1038/nmat4886} {\bibfield  {journal} {\bibinfo  {journal}
  {Nat. Mater.}\ }\textbf {\bibinfo {volume} {16}},\ \bibinfo {pages} {712}
  (\bibinfo {year} {2017})}\BibitemShut {NoStop}%
\bibitem [{\citenamefont {Cao}\ \emph {et~al.}(2020)\citenamefont {Cao},
  \citenamefont {Sheng}, \citenamefont {Edmonds}, \citenamefont {Ji},
  \citenamefont {Zheng},\ and\ \citenamefont {Wang}}]{admatCao2020}%
  \BibitemOpen
  \bibfield  {author} {\bibinfo {author} {\bibfnamefont {Y.}~\bibnamefont
  {Cao}}, \bibinfo {author} {\bibfnamefont {Y.}~\bibnamefont {Sheng}}, \bibinfo
  {author} {\bibfnamefont {K.~W.}\ \bibnamefont {Edmonds}}, \bibinfo {author}
  {\bibfnamefont {Y.}~\bibnamefont {Ji}}, \bibinfo {author} {\bibfnamefont
  {H.}~\bibnamefont {Zheng}}, \ and\ \bibinfo {author} {\bibfnamefont
  {K.}~\bibnamefont {Wang}},\ }\href {\doibase
  https://doi.org/10.1002/adma.201907929} {\bibfield  {journal} {\bibinfo
  {journal} {Adv. Mater.}\ }\textbf {\bibinfo {volume} {32}},\ \bibinfo {pages}
  {1907929} (\bibinfo {year} {2020})}\BibitemShut {NoStop}%
\bibitem [{\citenamefont {Bekele}\ \emph {et~al.}(2021)\citenamefont {Bekele},
  \citenamefont {Liu}, \citenamefont {Cao},\ and\ \citenamefont
  {Wang}}]{adeleBekele2021}%
  \BibitemOpen
  \bibfield  {author} {\bibinfo {author} {\bibfnamefont {Z.~A.}\ \bibnamefont
  {Bekele}}, \bibinfo {author} {\bibfnamefont {X.}~\bibnamefont {Liu}},
  \bibinfo {author} {\bibfnamefont {Y.}~\bibnamefont {Cao}}, \ and\ \bibinfo
  {author} {\bibfnamefont {K.}~\bibnamefont {Wang}},\ }\href {\doibase
  https://doi.org/10.1002/aelm.202000793} {\bibfield  {journal} {\bibinfo
  {journal} {Adv. Electron. Mater.}\ }\textbf {\bibinfo {volume} {7}},\
  \bibinfo {pages} {2000793} (\bibinfo {year} {2021})}\BibitemShut {NoStop}%
\bibitem [{\citenamefont {Filianina}\ \emph {et~al.}(2020)\citenamefont
  {Filianina}, \citenamefont {Hanke}, \citenamefont {Lee}, \citenamefont {Han},
  \citenamefont {Jaiswal}, \citenamefont {Rajan}, \citenamefont {Jakob},
  \citenamefont {Mokrousov},\ and\ \citenamefont {Kl\"aui}}]{PRLFilianina2020}%
  \BibitemOpen
  \bibfield  {author} {\bibinfo {author} {\bibfnamefont {M.}~\bibnamefont
  {Filianina}}, \bibinfo {author} {\bibfnamefont {J.-P.}\ \bibnamefont
  {Hanke}}, \bibinfo {author} {\bibfnamefont {K.}~\bibnamefont {Lee}}, \bibinfo
  {author} {\bibfnamefont {D.-S.}\ \bibnamefont {Han}}, \bibinfo {author}
  {\bibfnamefont {S.}~\bibnamefont {Jaiswal}}, \bibinfo {author} {\bibfnamefont
  {A.}~\bibnamefont {Rajan}}, \bibinfo {author} {\bibfnamefont
  {G.}~\bibnamefont {Jakob}}, \bibinfo {author} {\bibfnamefont
  {Y.}~\bibnamefont {Mokrousov}}, \ and\ \bibinfo {author} {\bibfnamefont
  {M.}~\bibnamefont {Kl\"aui}},\ }\href {\doibase
  10.1103/PhysRevLett.124.217701} {\bibfield  {journal} {\bibinfo  {journal}
  {Phys. Rev. Lett.}\ }\textbf {\bibinfo {volume} {124}},\ \bibinfo {pages}
  {217701} (\bibinfo {year} {2020})}\BibitemShut {NoStop}%
\bibitem [{\citenamefont {Chiba}\ \emph {et~al.}(2017)\citenamefont {Chiba},
  \citenamefont {Takahashi},\ and\ \citenamefont {Bauer}}]{Chiba_PRB_2017}%
  \BibitemOpen
  \bibfield  {author} {\bibinfo {author} {\bibfnamefont {T.}~\bibnamefont
  {Chiba}}, \bibinfo {author} {\bibfnamefont {S.}~\bibnamefont {Takahashi}}, \
  and\ \bibinfo {author} {\bibfnamefont {G.~E.~W.}\ \bibnamefont {Bauer}},\
  }\href {\doibase 10.1103/PhysRevB.95.094428} {\bibfield  {journal} {\bibinfo
  {journal} {Phys. Rev. B}\ }\textbf {\bibinfo {volume} {95}},\ \bibinfo
  {pages} {094428} (\bibinfo {year} {2017})}\BibitemShut {NoStop}%
\bibitem [{\citenamefont {Chiba}\ and\ \citenamefont
  {Komine}(2020)}]{Chiba-PRA2020}%
  \BibitemOpen
  \bibfield  {author} {\bibinfo {author} {\bibfnamefont {T.}~\bibnamefont
  {Chiba}}\ and\ \bibinfo {author} {\bibfnamefont {T.}~\bibnamefont {Komine}},\
  }\href {\doibase 10.1103/PhysRevApplied.14.034031} {\bibfield  {journal}
  {\bibinfo  {journal} {Phys. Rev. Appl.}\ }\textbf {\bibinfo {volume} {14}},\
  \bibinfo {pages} {034031} (\bibinfo {year} {2020})}\BibitemShut {NoStop}%
\bibitem [{\citenamefont {Chiba}\ \emph {et~al.}(2021)\citenamefont {Chiba},
  \citenamefont {Leon},\ and\ \citenamefont {Komine}}]{Chiba-APL2021}%
  \BibitemOpen
  \bibfield  {author} {\bibinfo {author} {\bibfnamefont {T.}~\bibnamefont
  {Chiba}}, \bibinfo {author} {\bibfnamefont {A.~O.}\ \bibnamefont {Leon}}, \
  and\ \bibinfo {author} {\bibfnamefont {T.}~\bibnamefont {Komine}},\ }\href
  {\doibase 10.1063/5.0046217} {\bibfield  {journal} {\bibinfo  {journal}
  {Appl. Phys. Lett.}\ }\textbf {\bibinfo {volume} {118}},\ \bibinfo {pages}
  {252402} (\bibinfo {year} {2021})}\BibitemShut {NoStop}%
\bibitem [{\citenamefont {Weisheit}\ \emph {et~al.}(2007)\citenamefont
  {Weisheit}, \citenamefont {Fähler}, \citenamefont {Marty}, \citenamefont
  {Souche}, \citenamefont {Poinsignon},\ and\ \citenamefont
  {Givord}}]{Weisheit_Science_2007}%
  \BibitemOpen
  \bibfield  {author} {\bibinfo {author} {\bibfnamefont {M.}~\bibnamefont
  {Weisheit}}, \bibinfo {author} {\bibfnamefont {S.}~\bibnamefont {Fähler}},
  \bibinfo {author} {\bibfnamefont {A.}~\bibnamefont {Marty}}, \bibinfo
  {author} {\bibfnamefont {Y.}~\bibnamefont {Souche}}, \bibinfo {author}
  {\bibfnamefont {C.}~\bibnamefont {Poinsignon}}, \ and\ \bibinfo {author}
  {\bibfnamefont {D.}~\bibnamefont {Givord}},\ }\href {\doibase
  10.1126/science.1136629} {\bibfield  {journal} {\bibinfo  {journal}
  {Science}\ }\textbf {\bibinfo {volume} {315}},\ \bibinfo {pages} {349}
  (\bibinfo {year} {2007})}\BibitemShut {NoStop}%
\bibitem [{\citenamefont {Tsujikawa}\ and\ \citenamefont
  {Oda}(2009)}]{Tsujikawa_PRL_2009}%
  \BibitemOpen
  \bibfield  {author} {\bibinfo {author} {\bibfnamefont {M.}~\bibnamefont
  {Tsujikawa}}\ and\ \bibinfo {author} {\bibfnamefont {T.}~\bibnamefont
  {Oda}},\ }\href {\doibase 10.1103/PhysRevLett.102.247203} {\bibfield
  {journal} {\bibinfo  {journal} {Phys. Rev. Lett.}\ }\textbf {\bibinfo
  {volume} {102}},\ \bibinfo {pages} {247203} (\bibinfo {year}
  {2009})}\BibitemShut {NoStop}%
\bibitem [{\citenamefont {Nakamura}\ \emph {et~al.}(2009)\citenamefont
  {Nakamura}, \citenamefont {Shimabukuro}, \citenamefont {Fujiwara},
  \citenamefont {Akiyama}, \citenamefont {Ito},\ and\ \citenamefont
  {Freeman}}]{Nakamura_PRL_2009}%
  \BibitemOpen
  \bibfield  {author} {\bibinfo {author} {\bibfnamefont {K.}~\bibnamefont
  {Nakamura}}, \bibinfo {author} {\bibfnamefont {R.}~\bibnamefont
  {Shimabukuro}}, \bibinfo {author} {\bibfnamefont {Y.}~\bibnamefont
  {Fujiwara}}, \bibinfo {author} {\bibfnamefont {T.}~\bibnamefont {Akiyama}},
  \bibinfo {author} {\bibfnamefont {T.}~\bibnamefont {Ito}}, \ and\ \bibinfo
  {author} {\bibfnamefont {A.~J.}\ \bibnamefont {Freeman}},\ }\href {\doibase
  10.1103/PhysRevLett.102.187201} {\bibfield  {journal} {\bibinfo  {journal}
  {Phys. Rev. Lett.}\ }\textbf {\bibinfo {volume} {102}},\ \bibinfo {pages}
  {187201} (\bibinfo {year} {2009})}\BibitemShut {NoStop}%
\bibitem [{\citenamefont {Duan}\ \emph {et~al.}(2008)\citenamefont {Duan},
  \citenamefont {Velev}, \citenamefont {Sabirianov}, \citenamefont {Zhu},
  \citenamefont {Chu}, \citenamefont {Jaswal},\ and\ \citenamefont
  {Tsymbal}}]{Duan_PRL_2008}%
  \BibitemOpen
  \bibfield  {author} {\bibinfo {author} {\bibfnamefont {C.-G.}\ \bibnamefont
  {Duan}}, \bibinfo {author} {\bibfnamefont {J.~P.}\ \bibnamefont {Velev}},
  \bibinfo {author} {\bibfnamefont {R.~F.}\ \bibnamefont {Sabirianov}},
  \bibinfo {author} {\bibfnamefont {Z.}~\bibnamefont {Zhu}}, \bibinfo {author}
  {\bibfnamefont {J.}~\bibnamefont {Chu}}, \bibinfo {author} {\bibfnamefont
  {S.~S.}\ \bibnamefont {Jaswal}}, \ and\ \bibinfo {author} {\bibfnamefont
  {E.~Y.}\ \bibnamefont {Tsymbal}},\ }\href {\doibase
  10.1103/PhysRevLett.101.137201} {\bibfield  {journal} {\bibinfo  {journal}
  {Phys. Rev. Lett.}\ }\textbf {\bibinfo {volume} {101}},\ \bibinfo {pages}
  {137201} (\bibinfo {year} {2008})}\BibitemShut {NoStop}%
\bibitem [{\citenamefont {Shiota}\ \emph {et~al.}(2009)\citenamefont {Shiota},
  \citenamefont {Maruyama}, \citenamefont {Nozaki}, \citenamefont {Shinjo},
  \citenamefont {Shiraishi},\ and\ \citenamefont {Suzuki}}]{Shiota_APEX_2009}%
  \BibitemOpen
  \bibfield  {author} {\bibinfo {author} {\bibfnamefont {Y.}~\bibnamefont
  {Shiota}}, \bibinfo {author} {\bibfnamefont {T.}~\bibnamefont {Maruyama}},
  \bibinfo {author} {\bibfnamefont {T.}~\bibnamefont {Nozaki}}, \bibinfo
  {author} {\bibfnamefont {T.}~\bibnamefont {Shinjo}}, \bibinfo {author}
  {\bibfnamefont {M.}~\bibnamefont {Shiraishi}}, \ and\ \bibinfo {author}
  {\bibfnamefont {Y.}~\bibnamefont {Suzuki}},\ }\href {\doibase
  10.1143/APEX.2.063001} {\bibfield  {journal} {\bibinfo  {journal} {Appl.
  Phys. Express}\ }\textbf {\bibinfo {volume} {2}},\ \bibinfo {pages} {063001}
  (\bibinfo {year} {2009})}\BibitemShut {NoStop}%
\bibitem [{\citenamefont {Khalili~Amiri}\ \emph {et~al.}(2013)\citenamefont
  {Khalili~Amiri}, \citenamefont {Upadhyaya}, \citenamefont {Alzate},\ and\
  \citenamefont {Wang}}]{Amiri_JAP_2013}%
  \BibitemOpen
  \bibfield  {author} {\bibinfo {author} {\bibfnamefont {P.}~\bibnamefont
  {Khalili~Amiri}}, \bibinfo {author} {\bibfnamefont {P.}~\bibnamefont
  {Upadhyaya}}, \bibinfo {author} {\bibfnamefont {J.~G.}\ \bibnamefont
  {Alzate}}, \ and\ \bibinfo {author} {\bibfnamefont {K.~L.}\ \bibnamefont
  {Wang}},\ }\href {\doibase 10.1063/1.4773342} {\bibfield  {journal} {\bibinfo
   {journal} {J. Appl. Phys.}\ }\textbf {\bibinfo {volume} {113}},\ \bibinfo
  {pages} {013912} (\bibinfo {year} {2013})}\BibitemShut {NoStop}%
\bibitem [{\citenamefont {Han}\ \emph {et~al.}(2015)\citenamefont {Han},
  \citenamefont {Meng}, \citenamefont {Huang}, \citenamefont {Naik},
  \citenamefont {Sim}, \citenamefont {Tran},\ and\ \citenamefont
  {Lim}}]{Han_IEEE_2015}%
  \BibitemOpen
  \bibfield  {author} {\bibinfo {author} {\bibfnamefont {G.}~\bibnamefont
  {Han}}, \bibinfo {author} {\bibfnamefont {H.}~\bibnamefont {Meng}}, \bibinfo
  {author} {\bibfnamefont {J.}~\bibnamefont {Huang}}, \bibinfo {author}
  {\bibfnamefont {V.}~\bibnamefont {Naik}}, \bibinfo {author} {\bibfnamefont
  {C.}~\bibnamefont {Sim}}, \bibinfo {author} {\bibfnamefont {M.}~\bibnamefont
  {Tran}}, \ and\ \bibinfo {author} {\bibfnamefont {T.}~\bibnamefont {Lim}},\
  }\href {\doibase 10.1109/TMAG.2014.2354452} {\bibfield  {journal} {\bibinfo
  {journal} {IEEE Trans. Magn.}\ }\textbf {\bibinfo {volume} {51}},\ \bibinfo
  {pages} {1} (\bibinfo {year} {2015})}\BibitemShut {NoStop}%
\bibitem [{\citenamefont {Kanai}\ \emph {et~al.}(2014)\citenamefont {Kanai},
  \citenamefont {Nakatani}, \citenamefont {Yamanouchi}, \citenamefont {Ikeda},
  \citenamefont {Sato}, \citenamefont {Matsukura},\ and\ \citenamefont
  {Ohno}}]{Kanai_APL_2014}%
  \BibitemOpen
  \bibfield  {author} {\bibinfo {author} {\bibfnamefont {S.}~\bibnamefont
  {Kanai}}, \bibinfo {author} {\bibfnamefont {Y.}~\bibnamefont {Nakatani}},
  \bibinfo {author} {\bibfnamefont {M.}~\bibnamefont {Yamanouchi}}, \bibinfo
  {author} {\bibfnamefont {S.}~\bibnamefont {Ikeda}}, \bibinfo {author}
  {\bibfnamefont {H.}~\bibnamefont {Sato}}, \bibinfo {author} {\bibfnamefont
  {F.}~\bibnamefont {Matsukura}}, \ and\ \bibinfo {author} {\bibfnamefont
  {H.}~\bibnamefont {Ohno}},\ }\href {\doibase 10.1063/1.4880720} {\bibfield
  {journal} {\bibinfo  {journal} {Appl. Phys. Lett.}\ }\textbf {\bibinfo
  {volume} {104}},\ \bibinfo {pages} {212406} (\bibinfo {year}
  {2014})}\BibitemShut {NoStop}%
\bibitem [{\citenamefont {Kato}\ \emph {et~al.}(2018)\citenamefont {Kato},
  \citenamefont {Yoda}, \citenamefont {Saito}, \citenamefont {Oikawa},
  \citenamefont {Fujii}, \citenamefont {Yoshiki}, \citenamefont {Koi},
  \citenamefont {Sugiyama}, \citenamefont {Ishikawa}, \citenamefont {Inokuchi},
  \citenamefont {Shimomura}, \citenamefont {Shimizu}, \citenamefont
  {Shirotori}, \citenamefont {Altansargai}, \citenamefont {Ohsawa},
  \citenamefont {Ikegami}, \citenamefont {Tiwari},\ and\ \citenamefont
  {Kurobe}}]{Kato_2018}%
  \BibitemOpen
  \bibfield  {author} {\bibinfo {author} {\bibfnamefont {Y.}~\bibnamefont
  {Kato}}, \bibinfo {author} {\bibfnamefont {H.}~\bibnamefont {Yoda}}, \bibinfo
  {author} {\bibfnamefont {Y.}~\bibnamefont {Saito}}, \bibinfo {author}
  {\bibfnamefont {S.}~\bibnamefont {Oikawa}}, \bibinfo {author} {\bibfnamefont
  {K.}~\bibnamefont {Fujii}}, \bibinfo {author} {\bibfnamefont
  {M.}~\bibnamefont {Yoshiki}}, \bibinfo {author} {\bibfnamefont
  {K.}~\bibnamefont {Koi}}, \bibinfo {author} {\bibfnamefont {H.}~\bibnamefont
  {Sugiyama}}, \bibinfo {author} {\bibfnamefont {M.}~\bibnamefont {Ishikawa}},
  \bibinfo {author} {\bibfnamefont {T.}~\bibnamefont {Inokuchi}}, \bibinfo
  {author} {\bibfnamefont {N.}~\bibnamefont {Shimomura}}, \bibinfo {author}
  {\bibfnamefont {M.}~\bibnamefont {Shimizu}}, \bibinfo {author} {\bibfnamefont
  {S.}~\bibnamefont {Shirotori}}, \bibinfo {author} {\bibfnamefont
  {B.}~\bibnamefont {Altansargai}}, \bibinfo {author} {\bibfnamefont
  {Y.}~\bibnamefont {Ohsawa}}, \bibinfo {author} {\bibfnamefont
  {K.}~\bibnamefont {Ikegami}}, \bibinfo {author} {\bibfnamefont
  {A.}~\bibnamefont {Tiwari}}, \ and\ \bibinfo {author} {\bibfnamefont
  {A.}~\bibnamefont {Kurobe}},\ }\href {\doibase 10.7567/APEX.11.053007}
  {\bibfield  {journal} {\bibinfo  {journal} {Appl. Phys. Express}\ }\textbf
  {\bibinfo {volume} {11}},\ \bibinfo {pages} {053007} (\bibinfo {year}
  {2018})}\BibitemShut {NoStop}%
\bibitem [{\citenamefont {Yamamoto}\ \emph {et~al.}(2018)\citenamefont
  {Yamamoto}, \citenamefont {Nozaki}, \citenamefont {Shiota}, \citenamefont
  {Imamura}, \citenamefont {Tamaru}, \citenamefont {Yakushiji}, \citenamefont
  {Kubota}, \citenamefont {Fukushima}, \citenamefont {Suzuki},\ and\
  \citenamefont {Yuasa}}]{Yamamoto_PRA_2018}%
  \BibitemOpen
  \bibfield  {author} {\bibinfo {author} {\bibfnamefont {T.}~\bibnamefont
  {Yamamoto}}, \bibinfo {author} {\bibfnamefont {T.}~\bibnamefont {Nozaki}},
  \bibinfo {author} {\bibfnamefont {Y.}~\bibnamefont {Shiota}}, \bibinfo
  {author} {\bibfnamefont {H.}~\bibnamefont {Imamura}}, \bibinfo {author}
  {\bibfnamefont {S.}~\bibnamefont {Tamaru}}, \bibinfo {author} {\bibfnamefont
  {K.}~\bibnamefont {Yakushiji}}, \bibinfo {author} {\bibfnamefont
  {H.}~\bibnamefont {Kubota}}, \bibinfo {author} {\bibfnamefont
  {A.}~\bibnamefont {Fukushima}}, \bibinfo {author} {\bibfnamefont
  {Y.}~\bibnamefont {Suzuki}}, \ and\ \bibinfo {author} {\bibfnamefont
  {S.}~\bibnamefont {Yuasa}},\ }\href {\doibase
  10.1103/PhysRevApplied.10.024004} {\bibfield  {journal} {\bibinfo  {journal}
  {Phys. Rev. Appl.}\ }\textbf {\bibinfo {volume} {10}},\ \bibinfo {pages}
  {024004} (\bibinfo {year} {2018})}\BibitemShut {NoStop}%
\bibitem [{\citenamefont {Yoda}\ \emph {et~al.}(2016)\citenamefont {Yoda},
  \citenamefont {Shimomura}, \citenamefont {Ohsawa}, \citenamefont {Shirotori},
  \citenamefont {Kato}, \citenamefont {Inokuchi}, \citenamefont {Kamiguchi},
  \citenamefont {Altansargai}, \citenamefont {Saito}, \citenamefont {Koi},
  \citenamefont {Sugiyama}, \citenamefont {Oikawa}, \citenamefont {Shimizu},
  \citenamefont {Ishikawa}, \citenamefont {Ikegami},\ and\ \citenamefont
  {Kurobe}}]{Yoda_IEDM_2016}%
  \BibitemOpen
  \bibfield  {author} {\bibinfo {author} {\bibfnamefont {H.}~\bibnamefont
  {Yoda}}, \bibinfo {author} {\bibfnamefont {N.}~\bibnamefont {Shimomura}},
  \bibinfo {author} {\bibfnamefont {Y.}~\bibnamefont {Ohsawa}}, \bibinfo
  {author} {\bibfnamefont {S.}~\bibnamefont {Shirotori}}, \bibinfo {author}
  {\bibfnamefont {Y.}~\bibnamefont {Kato}}, \bibinfo {author} {\bibfnamefont
  {T.}~\bibnamefont {Inokuchi}}, \bibinfo {author} {\bibfnamefont
  {Y.}~\bibnamefont {Kamiguchi}}, \bibinfo {author} {\bibfnamefont
  {B.}~\bibnamefont {Altansargai}}, \bibinfo {author} {\bibfnamefont
  {Y.}~\bibnamefont {Saito}}, \bibinfo {author} {\bibfnamefont
  {K.}~\bibnamefont {Koi}}, \bibinfo {author} {\bibfnamefont {H.}~\bibnamefont
  {Sugiyama}}, \bibinfo {author} {\bibfnamefont {S.}~\bibnamefont {Oikawa}},
  \bibinfo {author} {\bibfnamefont {M.}~\bibnamefont {Shimizu}}, \bibinfo
  {author} {\bibfnamefont {M.}~\bibnamefont {Ishikawa}}, \bibinfo {author}
  {\bibfnamefont {K.}~\bibnamefont {Ikegami}}, \ and\ \bibinfo {author}
  {\bibfnamefont {A.}~\bibnamefont {Kurobe}},\ }in\ \href {\doibase
  10.1109/IEDM.2016.7838495} {\emph {\bibinfo {booktitle} {2016 IEEE
  International Electron Devices Meeting (IEDM)}}}\ (\bibinfo {year} {2016})\
  pp.\ \bibinfo {pages} {27.6.1--27.6.4}\BibitemShut {NoStop}%
\bibitem [{\citenamefont {Inokuchi}\ \emph {et~al.}(2017)\citenamefont
  {Inokuchi}, \citenamefont {Yoda}, \citenamefont {Kato}, \citenamefont
  {Shimizu}, \citenamefont {Shirotori}, \citenamefont {Shimomura},
  \citenamefont {Koi}, \citenamefont {Kamiguchi}, \citenamefont {Sugiyama},
  \citenamefont {Oikawa}, \citenamefont {Ikegami}, \citenamefont {Ishikawa},
  \citenamefont {Altansargai}, \citenamefont {Tiwari}, \citenamefont {Ohsawa},
  \citenamefont {Saito},\ and\ \citenamefont {Kurobe}}]{Inokuchi_2017}%
  \BibitemOpen
  \bibfield  {author} {\bibinfo {author} {\bibfnamefont {T.}~\bibnamefont
  {Inokuchi}}, \bibinfo {author} {\bibfnamefont {H.}~\bibnamefont {Yoda}},
  \bibinfo {author} {\bibfnamefont {Y.}~\bibnamefont {Kato}}, \bibinfo {author}
  {\bibfnamefont {M.}~\bibnamefont {Shimizu}}, \bibinfo {author} {\bibfnamefont
  {S.}~\bibnamefont {Shirotori}}, \bibinfo {author} {\bibfnamefont
  {N.}~\bibnamefont {Shimomura}}, \bibinfo {author} {\bibfnamefont
  {K.}~\bibnamefont {Koi}}, \bibinfo {author} {\bibfnamefont {Y.}~\bibnamefont
  {Kamiguchi}}, \bibinfo {author} {\bibfnamefont {H.}~\bibnamefont {Sugiyama}},
  \bibinfo {author} {\bibfnamefont {S.}~\bibnamefont {Oikawa}}, \bibinfo
  {author} {\bibfnamefont {K.}~\bibnamefont {Ikegami}}, \bibinfo {author}
  {\bibfnamefont {M.}~\bibnamefont {Ishikawa}}, \bibinfo {author}
  {\bibfnamefont {B.}~\bibnamefont {Altansargai}}, \bibinfo {author}
  {\bibfnamefont {A.}~\bibnamefont {Tiwari}}, \bibinfo {author} {\bibfnamefont
  {Y.}~\bibnamefont {Ohsawa}}, \bibinfo {author} {\bibfnamefont
  {Y.}~\bibnamefont {Saito}}, \ and\ \bibinfo {author} {\bibfnamefont
  {A.}~\bibnamefont {Kurobe}},\ }\href {\doibase 10.1063/1.4986923} {\bibfield
  {journal} {\bibinfo  {journal} {Appl. Phys. Lett.}\ }\textbf {\bibinfo
  {volume} {110}},\ \bibinfo {pages} {252404} (\bibinfo {year}
  {2017})}\BibitemShut {NoStop}%
\bibitem [{\citenamefont {Baek}\ \emph {et~al.}(2018)\citenamefont {Baek},
  \citenamefont {Park}, \citenamefont {Kil}, \citenamefont {Jang},
  \citenamefont {Park}, \citenamefont {Lee},\ and\ \citenamefont
  {Park}}]{Baek_NatElec_2018}%
  \BibitemOpen
  \bibfield  {author} {\bibinfo {author} {\bibfnamefont {S.-h.~C.}\
  \bibnamefont {Baek}}, \bibinfo {author} {\bibfnamefont {K.-W.}\ \bibnamefont
  {Park}}, \bibinfo {author} {\bibfnamefont {D.-S.}\ \bibnamefont {Kil}},
  \bibinfo {author} {\bibfnamefont {Y.}~\bibnamefont {Jang}}, \bibinfo {author}
  {\bibfnamefont {J.}~\bibnamefont {Park}}, \bibinfo {author} {\bibfnamefont
  {K.-J.}\ \bibnamefont {Lee}}, \ and\ \bibinfo {author} {\bibfnamefont
  {B.-G.}\ \bibnamefont {Park}},\ }\href {\doibase 10.1038/s41928-018-0099-8}
  {\bibfield  {journal} {\bibinfo  {journal} {Nat. Electron.}\ }\textbf
  {\bibinfo {volume} {1}},\ \bibinfo {pages} {398} (\bibinfo {year}
  {2018})}\BibitemShut {NoStop}%
\bibitem [{\citenamefont {Mishra}\ \emph {et~al.}(2019)\citenamefont {Mishra},
  \citenamefont {Mahfouzi}, \citenamefont {Kumar}, \citenamefont {Cai},
  \citenamefont {Chen}, \citenamefont {Qiu}, \citenamefont {Kioussis},\ and\
  \citenamefont {Yang}}]{Mishra_NatComm_2019}%
  \BibitemOpen
  \bibfield  {author} {\bibinfo {author} {\bibfnamefont {R.}~\bibnamefont
  {Mishra}}, \bibinfo {author} {\bibfnamefont {F.}~\bibnamefont {Mahfouzi}},
  \bibinfo {author} {\bibfnamefont {D.}~\bibnamefont {Kumar}}, \bibinfo
  {author} {\bibfnamefont {K.}~\bibnamefont {Cai}}, \bibinfo {author}
  {\bibfnamefont {M.}~\bibnamefont {Chen}}, \bibinfo {author} {\bibfnamefont
  {X.}~\bibnamefont {Qiu}}, \bibinfo {author} {\bibfnamefont {N.}~\bibnamefont
  {Kioussis}}, \ and\ \bibinfo {author} {\bibfnamefont {H.}~\bibnamefont
  {Yang}},\ }\href {\doibase 10.1038/s41467-018-08274-8} {\bibfield  {journal}
  {\bibinfo  {journal} {Nat. Commun.}\ }\textbf {\bibinfo {volume} {10}},\
  \bibinfo {pages} {248} (\bibinfo {year} {2019})}\BibitemShut {NoStop}%
\bibitem [{\citenamefont {Hirahara}\ \emph {et~al.}(2017)\citenamefont
  {Hirahara}, \citenamefont {Eremeev}, \citenamefont {Shirasawa}, \citenamefont
  {Okuyama}, \citenamefont {Kubo}, \citenamefont {Nakanishi}, \citenamefont
  {Akiyama}, \citenamefont {Takayama}, \citenamefont {Hajiri}, \citenamefont
  {Ideta}, \citenamefont {Matsunami}, \citenamefont {Sumida}, \citenamefont
  {Miyamoto}, \citenamefont {Takagi}, \citenamefont {Tanaka}, \citenamefont
  {Okuda}, \citenamefont {Yokoyama}, \citenamefont {Kimura}, \citenamefont
  {Hasegawa},\ and\ \citenamefont {Chulkov}}]{Hirahara_NanoLett_2017}%
  \BibitemOpen
  \bibfield  {author} {\bibinfo {author} {\bibfnamefont {T.}~\bibnamefont
  {Hirahara}}, \bibinfo {author} {\bibfnamefont {S.~V.}\ \bibnamefont
  {Eremeev}}, \bibinfo {author} {\bibfnamefont {T.}~\bibnamefont {Shirasawa}},
  \bibinfo {author} {\bibfnamefont {Y.}~\bibnamefont {Okuyama}}, \bibinfo
  {author} {\bibfnamefont {T.}~\bibnamefont {Kubo}}, \bibinfo {author}
  {\bibfnamefont {R.}~\bibnamefont {Nakanishi}}, \bibinfo {author}
  {\bibfnamefont {R.}~\bibnamefont {Akiyama}}, \bibinfo {author} {\bibfnamefont
  {A.}~\bibnamefont {Takayama}}, \bibinfo {author} {\bibfnamefont
  {T.}~\bibnamefont {Hajiri}}, \bibinfo {author} {\bibfnamefont {S.-i.}\
  \bibnamefont {Ideta}}, \bibinfo {author} {\bibfnamefont {M.}~\bibnamefont
  {Matsunami}}, \bibinfo {author} {\bibfnamefont {K.}~\bibnamefont {Sumida}},
  \bibinfo {author} {\bibfnamefont {K.}~\bibnamefont {Miyamoto}}, \bibinfo
  {author} {\bibfnamefont {Y.}~\bibnamefont {Takagi}}, \bibinfo {author}
  {\bibfnamefont {K.}~\bibnamefont {Tanaka}}, \bibinfo {author} {\bibfnamefont
  {T.}~\bibnamefont {Okuda}}, \bibinfo {author} {\bibfnamefont
  {T.}~\bibnamefont {Yokoyama}}, \bibinfo {author} {\bibfnamefont {S.-i.}\
  \bibnamefont {Kimura}}, \bibinfo {author} {\bibfnamefont {S.}~\bibnamefont
  {Hasegawa}}, \ and\ \bibinfo {author} {\bibfnamefont {E.~V.}\ \bibnamefont
  {Chulkov}},\ }\href {\doibase 10.1021/acs.nanolett.7b00560} {\bibfield
  {journal} {\bibinfo  {journal} {Nano Letters}\ }\textbf {\bibinfo {volume}
  {17}},\ \bibinfo {pages} {3493} (\bibinfo {year} {2017})}\BibitemShut
  {NoStop}%
\bibitem [{\citenamefont {Mogi}\ \emph {et~al.}(2019)\citenamefont {Mogi},
  \citenamefont {Nakajima}, \citenamefont {Ukleev}, \citenamefont {Tsukazaki},
  \citenamefont {Yoshimi}, \citenamefont {Kawamura}, \citenamefont {Takahashi},
  \citenamefont {Hanashima}, \citenamefont {Kakurai}, \citenamefont {Arima},
  \citenamefont {Kawasaki},\ and\ \citenamefont {Tokura}}]{Mogi_PRL_2019}%
  \BibitemOpen
  \bibfield  {author} {\bibinfo {author} {\bibfnamefont {M.}~\bibnamefont
  {Mogi}}, \bibinfo {author} {\bibfnamefont {T.}~\bibnamefont {Nakajima}},
  \bibinfo {author} {\bibfnamefont {V.}~\bibnamefont {Ukleev}}, \bibinfo
  {author} {\bibfnamefont {A.}~\bibnamefont {Tsukazaki}}, \bibinfo {author}
  {\bibfnamefont {R.}~\bibnamefont {Yoshimi}}, \bibinfo {author} {\bibfnamefont
  {M.}~\bibnamefont {Kawamura}}, \bibinfo {author} {\bibfnamefont {K.~S.}\
  \bibnamefont {Takahashi}}, \bibinfo {author} {\bibfnamefont {T.}~\bibnamefont
  {Hanashima}}, \bibinfo {author} {\bibfnamefont {K.}~\bibnamefont {Kakurai}},
  \bibinfo {author} {\bibfnamefont {T.-h.}\ \bibnamefont {Arima}}, \bibinfo
  {author} {\bibfnamefont {M.}~\bibnamefont {Kawasaki}}, \ and\ \bibinfo
  {author} {\bibfnamefont {Y.}~\bibnamefont {Tokura}},\ }\href {\doibase
  10.1103/PhysRevLett.123.016804} {\bibfield  {journal} {\bibinfo  {journal}
  {Phys. Rev. Lett.}\ }\textbf {\bibinfo {volume} {123}},\ \bibinfo {pages}
  {016804} (\bibinfo {year} {2019})}\BibitemShut {NoStop}%
\bibitem [{\citenamefont {Fan}\ \emph {et~al.}(2016{\natexlab{b}})\citenamefont
  {Fan}, \citenamefont {Kou}, \citenamefont {Upadhyaya}, \citenamefont {Shao},
  \citenamefont {Pan}, \citenamefont {Lang}, \citenamefont {Che}, \citenamefont
  {Tang}, \citenamefont {Montazeri}, \citenamefont {Murata}, \citenamefont
  {Chang}, \citenamefont {Akyol}, \citenamefont {Yu}, \citenamefont {Nie},
  \citenamefont {Wong}, \citenamefont {Liu}, \citenamefont {Wang},
  \citenamefont {Tserkovnyak},\ and\ \citenamefont {Wang}}]{Fan_NatNano_2016}%
  \BibitemOpen
  \bibfield  {author} {\bibinfo {author} {\bibfnamefont {Y.}~\bibnamefont
  {Fan}}, \bibinfo {author} {\bibfnamefont {X.}~\bibnamefont {Kou}}, \bibinfo
  {author} {\bibfnamefont {P.}~\bibnamefont {Upadhyaya}}, \bibinfo {author}
  {\bibfnamefont {Q.}~\bibnamefont {Shao}}, \bibinfo {author} {\bibfnamefont
  {L.}~\bibnamefont {Pan}}, \bibinfo {author} {\bibfnamefont {M.}~\bibnamefont
  {Lang}}, \bibinfo {author} {\bibfnamefont {X.}~\bibnamefont {Che}}, \bibinfo
  {author} {\bibfnamefont {J.}~\bibnamefont {Tang}}, \bibinfo {author}
  {\bibfnamefont {M.}~\bibnamefont {Montazeri}}, \bibinfo {author}
  {\bibfnamefont {K.}~\bibnamefont {Murata}}, \bibinfo {author} {\bibfnamefont
  {L.-T.}\ \bibnamefont {Chang}}, \bibinfo {author} {\bibfnamefont
  {M.}~\bibnamefont {Akyol}}, \bibinfo {author} {\bibfnamefont
  {G.}~\bibnamefont {Yu}}, \bibinfo {author} {\bibfnamefont {T.}~\bibnamefont
  {Nie}}, \bibinfo {author} {\bibfnamefont {K.~L.}\ \bibnamefont {Wong}},
  \bibinfo {author} {\bibfnamefont {J.}~\bibnamefont {Liu}}, \bibinfo {author}
  {\bibfnamefont {Y.}~\bibnamefont {Wang}}, \bibinfo {author} {\bibfnamefont
  {Y.}~\bibnamefont {Tserkovnyak}}, \ and\ \bibinfo {author} {\bibfnamefont
  {K.~L.}\ \bibnamefont {Wang}},\ }\href {\doibase 10.1038/nnano.2015.294}
  {\bibfield  {journal} {\bibinfo  {journal} {Nat. Nanotech.}\ }\textbf
  {\bibinfo {volume} {11}},\ \bibinfo {pages} {352} (\bibinfo {year}
  {2016}{\natexlab{b}})}\BibitemShut {NoStop}%
\bibitem [{\citenamefont {Li}\ \emph {et~al.}(2013)\citenamefont {Li},
  \citenamefont {Fan}, \citenamefont {Ji}, \citenamefont {Liu}, \citenamefont
  {Pan},\ and\ \citenamefont {Qiao}}]{Li_PhysLettA_2013}%
  \BibitemOpen
  \bibfield  {author} {\bibinfo {author} {\bibfnamefont {B.}~\bibnamefont
  {Li}}, \bibinfo {author} {\bibfnamefont {Q.}~\bibnamefont {Fan}}, \bibinfo
  {author} {\bibfnamefont {F.}~\bibnamefont {Ji}}, \bibinfo {author}
  {\bibfnamefont {Z.}~\bibnamefont {Liu}}, \bibinfo {author} {\bibfnamefont
  {H.}~\bibnamefont {Pan}}, \ and\ \bibinfo {author} {\bibfnamefont
  {S.}~\bibnamefont {Qiao}},\ }\href {\doibase 10.1016/j.physleta.2013.05.020}
  {\bibfield  {journal} {\bibinfo  {journal} {Phys. Lett. A}\ }\textbf
  {\bibinfo {volume} {377}},\ \bibinfo {pages} {1925} (\bibinfo {year}
  {2013})}\BibitemShut {NoStop}%
\bibitem [{\citenamefont {Chang}\ \emph {et~al.}(2013)\citenamefont {Chang},
  \citenamefont {Zhang}, \citenamefont {Liu}, \citenamefont {Zhang},
  \citenamefont {Feng}, \citenamefont {Li}, \citenamefont {Wang}, \citenamefont
  {Chen}, \citenamefont {Dai}, \citenamefont {Fang}, \citenamefont {Qi},
  \citenamefont {Zhang}, \citenamefont {Wang}, \citenamefont {He},
  \citenamefont {Ma},\ and\ \citenamefont {Xue}}]{Chang_AdvMater_2013}%
  \BibitemOpen
  \bibfield  {author} {\bibinfo {author} {\bibfnamefont {C.-Z.}\ \bibnamefont
  {Chang}}, \bibinfo {author} {\bibfnamefont {J.}~\bibnamefont {Zhang}},
  \bibinfo {author} {\bibfnamefont {M.}~\bibnamefont {Liu}}, \bibinfo {author}
  {\bibfnamefont {Z.}~\bibnamefont {Zhang}}, \bibinfo {author} {\bibfnamefont
  {X.}~\bibnamefont {Feng}}, \bibinfo {author} {\bibfnamefont {K.}~\bibnamefont
  {Li}}, \bibinfo {author} {\bibfnamefont {L.-L.}\ \bibnamefont {Wang}},
  \bibinfo {author} {\bibfnamefont {X.}~\bibnamefont {Chen}}, \bibinfo {author}
  {\bibfnamefont {X.}~\bibnamefont {Dai}}, \bibinfo {author} {\bibfnamefont
  {Z.}~\bibnamefont {Fang}}, \bibinfo {author} {\bibfnamefont {X.-L.}\
  \bibnamefont {Qi}}, \bibinfo {author} {\bibfnamefont {S.-C.}\ \bibnamefont
  {Zhang}}, \bibinfo {author} {\bibfnamefont {Y.}~\bibnamefont {Wang}},
  \bibinfo {author} {\bibfnamefont {K.}~\bibnamefont {He}}, \bibinfo {author}
  {\bibfnamefont {X.-C.}\ \bibnamefont {Ma}}, \ and\ \bibinfo {author}
  {\bibfnamefont {Q.-K.}\ \bibnamefont {Xue}},\ }\href {\doibase
  https://doi.org/10.1002/adma.201203493} {\bibfield  {journal} {\bibinfo
  {journal} {Adv. Mater.}\ }\textbf {\bibinfo {volume} {25}},\ \bibinfo {pages}
  {1065} (\bibinfo {year} {2013})}\BibitemShut {NoStop}%
\end{thebibliography}%

\end{document}